\documentclass[twocolumn,amsmath,amssymb,prb]{revtex4-2}
\usepackage{graphicx}
\usepackage{dcolumn}
\usepackage{bm}
\usepackage{braket}
\usepackage{color}

\begin{document}


\title{Pressure-induced phase transition in the $J_1$-$J_2$ square lattice antiferromagnet RbMoOPO$_4$Cl}




\author{Hikaru Takeda}
\affiliation{Institute for Solid State Physics, University of Tokyo, 5-1-5 Kashiwanoha, Kashiwa, Chiba 277-8581, Japan}%

\author{Touru Yamauchi}%
\affiliation{Institute for Solid State Physics, University of Tokyo, 5-1-5 Kashiwanoha, Kashiwa, Chiba 277-8581, Japan}%

\author{Masashi Takigawa}%
\affiliation{Institute for Solid State Physics, University of Tokyo, 5-1-5 Kashiwanoha, Kashiwa, Chiba 277-8581, Japan}%

\author{Hajime Ishikawa}%
\affiliation{Institute for Solid State Physics, University of Tokyo, 5-1-5 Kashiwanoha, Kashiwa, Chiba 277-8581, Japan}%

\author{Zenji Hiroi}%
\affiliation{Institute for Solid State Physics, University of Tokyo, 5-1-5 Kashiwanoha, Kashiwa, Chiba 277-8581, Japan}%

\date{\today}


\begin{abstract}
We report results of magnetization and $^{31}$P NMR measurements under high pressure up to 6.4~GPa on RbMoOPO$_4$Cl, which is a frustrated square-lattice antiferromagnet with competing nearest-neighbor and next-nearest-neighbor interactions. Anomalies in the pressure dependences of the NMR shift and the transferred hyperfine coupling constants indicate a structural phase transition at 2.6~GPa, which is likely to break mirror symmetry and triggers significant change of the exchange interactions. In fact, the NMR spectra in magnetically ordered states reveal a change from the columnar antiferromagnetic (CAF) order below 3.3~GPa to the N\'{e}el antiferromagnetic (NAF) order above 3.9~GPa. The spin lattice relaxation rate $1/T_1$ also indicates a change of dominant magnetic fluctuations from CAF-type to NAF-type with pressure. Although the NMR spectra in the intermediate pressure region  between 3.3 and 3.9 GPa show coexistence of the CAF and NAF phases, a certain component of $1/T_1$ shows paramagnetic behavior with persistent spin fluctuations, leaving possibility for a quantum disordered phase. The easy-plane anisotropy of spin fluctuations with unusual nonmonotonic temperature dependence at ambient pressure gets reversed to the Ising anisotropy at high pressures. This unexpected anisotropic behavior for a spin 1/2 system may be ascribed to the strong spin-orbit coupling of Mo-4$d$ electrons.        
\end{abstract}


\maketitle

\section{introduction}
For the past decades, great efforts have been devoted to discover novel quantum disordered states in frustrated spin systems~\cite{Lacroix}. One of the key concepts is the geometrical frustration inherent in triangle-based structures such as triangular, Kagome and pyrochlore lattices. In these lattices, exchange interactions on equivalent bonds compete strongly to destabilize magnetic order and may lead to quantum disordered ground states such as spin liquid or spin ice~\cite{Balents, Savary}.  Another type of frustration can be caused by competition between inequivalent exchange interactions, for example, those on the nearest-neighbor (NN) and the next-nearest-neighbor (NNN) bonds. Theories on such models have predicted, in addition to the quantum disordered phases, unconventional symmetry breaking with novel order parameters such as spin chirality, spin density wave, and spin nematic order~\cite{Hikihara, McCulloch, Kecke}. Experimentally, by taking advantage of chemical controllability of competing interactions, various frustrated antiferromagnets with quasi-one-dimensional~\cite{Nawa_LCVO1, Buttgen1, Buttgen2, Nawa_LCVO2, Orlova, Hase, Matsui, Dutton, Bosiocic, Grafe, Nawa_Na1, Nawa_Na2} or two-dimensional~\cite{Melzi1, Melzi2, Carretta1, Bombardi, Nath, Nath2, Carretta2, Tsirlin1, Tsirlin2, Tsirlin3, Bossoni, Ishikawa, Lezama, Yang} structures have been investigated.

\begin{figure}[b]
\includegraphics[width=7.5cm]{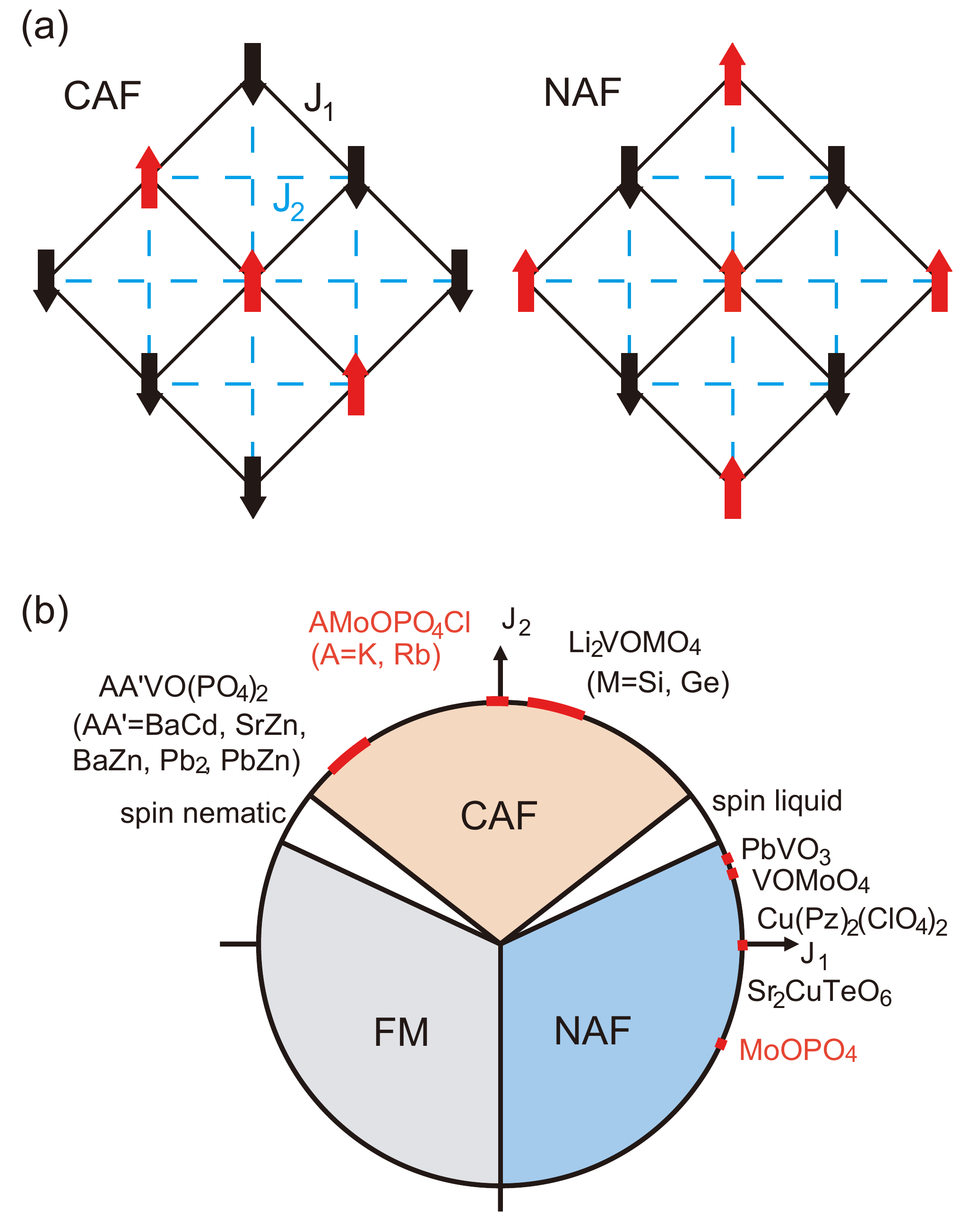}
\caption{\label{fig:J1J2} (Color online) (a) The columnar antiferromagnetic (CAF) and N\'{e}el antiferromagnetic (NAF) spin structure on the frustrated square lattice with competing exchange interactions $J_1$ and $J_2$ along the solid and dashed lines, respectively. (b) Schematic phase diagram of the spin-1/2 $J_1$-$J_2$ Heisenberg model on a square lattice with the proposed locations of candidate materials \cite{Yang}. The FM refers to the ferromagnetic state.} 
\end{figure}

The spin-1/2 Heisenberg model on a square lattice described by the Hamiltonian,
\begin{eqnarray}{\label{J1J2}}
{\mathcal{H}}= J_1\sum_{\rm{NN}}{\bf{S}}_i\cdot{\bf{S}}_j + J_2\sum_{\rm{NNN}}{\bf{S}}_i\cdot{\bf{S}}_k,
\end{eqnarray}
is a prototype of intensively studied frustrated systems of the second type~\cite{Misguich}.  As shown in Fig.~\ref{fig:J1J2}(a), the NN interaction $J_1$ along the side of the square competes with the NNN interaction $J_2$ along the diagonal when $J_2$ is antiferromagnetic ($J_2>0$). The phase diagram shown in Fig.~\ref{fig:J1J2}(b) contains three ordered phases: the columnar antiferromagnetic (CAF) phase stabilized for $J_2 \gtrsim |J_1|/2$, the N\'{e}el antiferromagnetic (NAF) phase for $J_2 \lesssim |J_1|/2$ with $J_1>0$, and ferromagnetic (FM) phase for $J_2 \lesssim |J_1|/2$ with $J_1<0$.  Theories have discussed the presence of a spin liquid or a spin nematic phase near the boundaries between CAF and NAF or FM phases~\cite{Sushkov, Shannon, Zhang, Capriotti1, Singh, Sirker, Capriotti2, Jiang, Mezzacapo, Gong, Shannon_nematic, Richter}.  Although several vanadium and molybdenum compounds have been synthesized, all materials investigated so far exhibit either NAF or CAF type magnetic order~\cite{Melzi1, Melzi2, Carretta1, Bombardi, Nath, Nath2, Carretta2, Tsirlin1, Tsirlin2, Tsirlin3, Bossoni, Ishikawa, Lezama, Yang}.  Since exchange interactions depend sensitively on structural parameters, application of high pressure is a promising root to discover novel quantum phases.  For example, ab-initio calculation of the exchange parameters in Li$_2$VOSiO$_4$ based on the structural data under high pressure predicts that application of 7.6 GPa of pressure would reduce the ratio  $J_2/J_1$ by 40$\%$~\cite{Pavarini}.  

\begin{figure}[b]
\includegraphics[width=8cm]{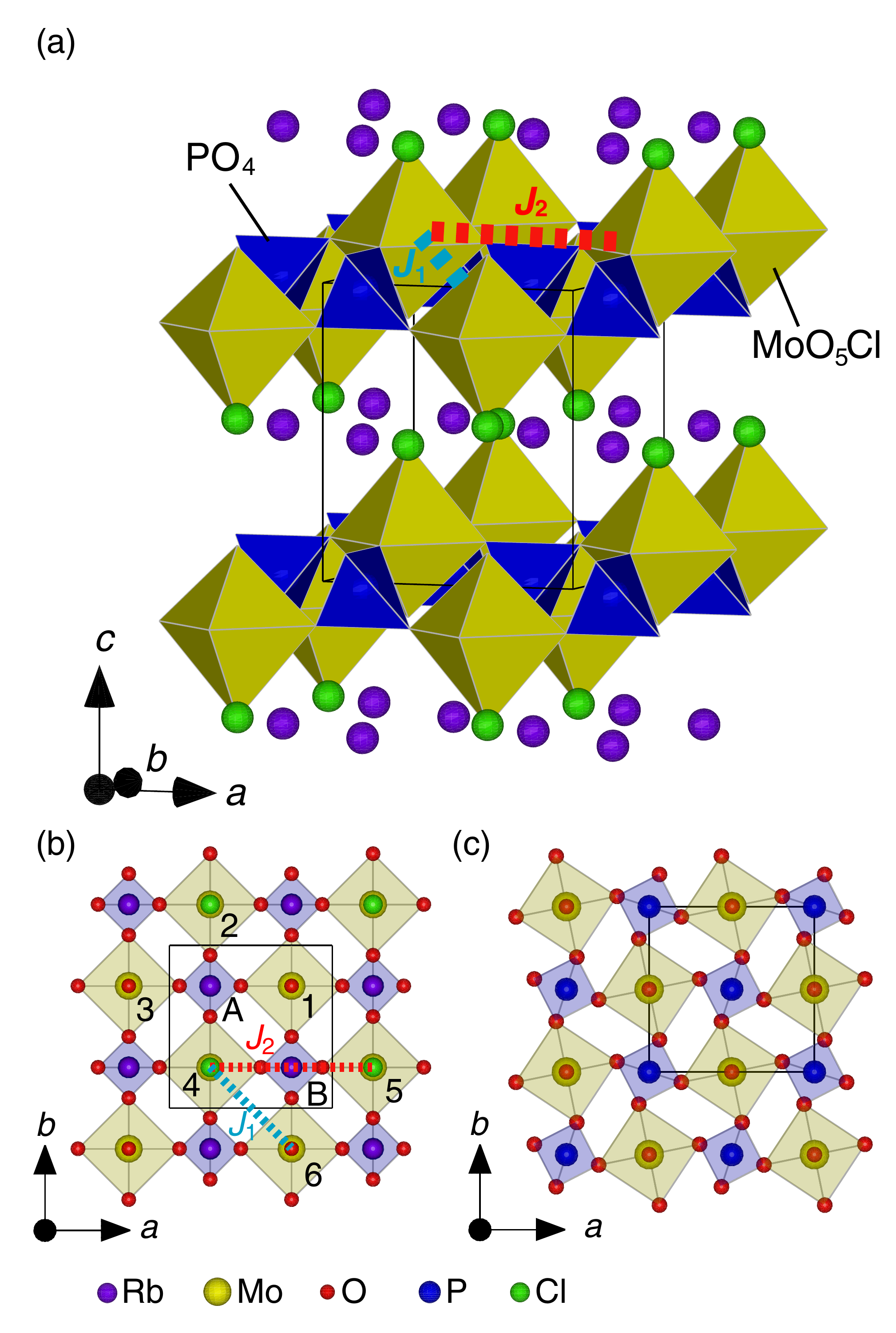}
\caption{\label{fig:RbMoOPO4Cl_MoOPO4} (Color online) (a) Perspective view of the crystal structure of RbMoOPO$_4$Cl and (b) projection along the $c$ axis.  The nearest-neighbor $J_1$ and next-nearest-neighbor $J_2$ exchange interactions are depicted by the blue and red dashed lines.  (c) The crystal structure of MoOPO$_4$ viewed along the $c$ axis.  Solid lines in (a)-(c) represent the unit cell.  These figures are drawn by VESTA~\cite{Momma}.} 
\end{figure}

Recently, Ishikawa $et$ $al$. reported magnetic properties of the square-lattice antiferromagnet RbMoOPO$_4$Cl containing Mo$^{5+}$ ions with spin 1/2~\cite{Ishikawa}. The crystal structure with the space group $P4/nmm$ consists of two-dimensional layers of corner-shared MoO$_5$Cl and PO$_4$ polyhedra separated by Rb layers, as shown in Figs. \ref{fig:RbMoOPO4Cl_MoOPO4}(a) and \ref{fig:RbMoOPO4Cl_MoOPO4}(b).  Each Mo layer consists of two Mo sheets at different heights along the $c$ axis.  Neighboring Mo atoms within a sheet are linearly connected via one PO$_4$ tetrahedron, while those at different sheets are connected at right angles via two PO$_4$ tetrahedra. The former and the latter connections generate $J_2$ and $J_1$, respectively. The effective moment obtained from the susceptibility data is 1.67 $\mu_{\rm{B}}$, which is close to 1.73 $\mu_{\rm{B}}$ for free spin 1/2~\cite{Ishikawa}, indicating the nondegenerate orbital state likely to be $d_{xy}$ type as is the case for other molybdenum and vanadium compounds~\cite{Melzi2, Carretta1, Tsirlin1, Tsirlin2, Tsirlin3,  Yang}. 

The temperature dependence of the susceptibility is fitted well to the high temperature series expansion of spin-1/2 square lattice Heisenberg model with only one exchange parameter $J=$ 29~K~\cite{Ishikawa, Roshbrooke}. No improvement of fitting was achieved by using the $J_1$-$J_2$ model, indicating that either $J_1$ or $J_2$ is very small.  Since the CAF order is confirmed below $T_{\rm{N}}$ = 8 K by the $^{31}$P-NMR and neutron scattering measurements~\cite{Ishikawa}, it is concluded that $J_2$ = 29 K and $J_1$ is negligibly small.  

When the Rb ions are replaced by K ions, the exchange constants changes as $J_1$ =$-2$ K and $J_2$ =19 K~\cite{Ishikawa}.  On the other hand, de-intercalation of the Rb and Cl ions generates another layered material MoOPO$_4$ with the space group $P4/n$, which orders in the NAF structure below $T_{\rm{N}}$ = 16 K~\cite{Yang}. The fitting of the susceptibility data gives $J_1$ = 11.4 K and $J_2$ = $-5.2$ K, corresponding to $J_2/J_1=-0.46$.  The large difference of the exchange interactions between RbMoOPO$_4$Cl and MoOPO$_4$ may be attributed to the different crystal symmetry~\cite{Yang}.  Unlike RbMoOPO$_4$Cl, the MoO$_6$ and PO$_4$ polyhedra in MoOPO$_4$ are twisted around the $c$-axis in opposite directions as shown in Fig.~\ref{fig:RbMoOPO4Cl_MoOPO4}(c).  This structural distortion is likely to cause significant changes in the hybridization between Mo-$4d$ and O-2$p$ orbitals, therefore, in the exchange constants.  

In this paper, we report results of magnetization and $^{31}$P NMR measurements on RbMoOPO$_4$Cl under high pressures.  Both the magnetization and the NMR frequency shift increase substantially with pressure $P$, indicating reduction of antiferromagnetic exchange interaction. The pressure dependence of the NMR shift shows a clear change of slope at $P_c$ = 2.6~GPa over a wide range of temperature, indicating a structural transition. This transition is likely to break the mirror symmetry with respect to the $ac$- and $bc$-planes owing to the twist of MoO$_5$Cl and PO$_4$ polyhedra similar to what was observed in MoOPO$_4$.  

The pressure dependence of the NMR spectra at low temperatures demonstrates the CAF magnetic order below 3.3~GPa and the NAF order above 3.9~GPa. The temperature dependence of the spin lattice relaxation rate $1/T_1$ also indicates a change of dominant magnetic fluctuations from CAF-type to NAF-type.  We argue that the drastic change of the ratio $J_2/J_1$, which is required to account for the CAF to NAF magnetic transition, is primarily driven by the change of structural symmetry above $P_c$.  
In the intermediate pressure region between 3.3~GPa and 3.9~GPa, the NMR spectra indicate coexistence of the CAF and NAF structures, pointing to a first-order transition with inhomogeneous distribution of the critical pressure. However, the nuclear relaxation curve at $P$ = 3.66~GPa reveals a component of $1/T_1$ showing paramagnetic behavior with persistent spin fluctuations down to the lowest temperature, unlike the behavior of either CAF or NAF phases. Thus a quantum disordered ground state in the intermediate pressure range still remains a possibility.  

Although unexpected for a spin 1/2 system, magnetic anisotropy with anomalous temperature and pressure dependences is observed.  At ambient pressure, both the CAF ordered moment and the spin fluctuations show easy plane anisotropy. Furthermore, the anisotropy of spin fluctuations exhibits anomalous nonmonotonic temperature dependence. The anisotropy gets reversed at higher pressures  where the NAF order is stabilized; both the NAF ordered moment and the critical spin fluctuations near $T_{\rm{N}}$ show strong Ising anisotropy. The puzzling anisotropic behavior appears to be a distinct feature of RbMoOPO$_4$Cl, presumably caused by strong spin orbit interaction of Mo-$4d$ electrons. 

\section{Experimental Procedure}
Polycrystalline samples prepared by the solid state reaction method were used for magnetization measurements~\cite{Ishikawa}. For NMR measurements, single crystals were grown by the flux method~\cite{Ishikawa}.  

Magnetization measurements were performed by using a SQUID magnetometer (Quantum Design, MPMS).  We used a piston-cylinder pressure cell at low pressures up to 1.6 GPa. A piece of Pb metal and 8.8~mg of the polycrystalline sample of RbMOPO$_4$Cl were placed in the pressure cell filled with glycerol as the pressure transmitting medium.  The pressure values were calibrated $in$ $situ$ from the known pressure dependence of the superconducting transition temperature of Pb metal~\cite{Eiling}.  

Magnetization measurements above 2 GPa were carried out by using an opposed-anvil pressure cell~\cite{Tateiwa, Yamauchi}.  A small amount of polycrystalline sample of 0.4~mg and Si-grease were placed in the cell.  As $in$ $situ$ pressure calibration was not possible because of the limited sample volume, pressure values were determined from separate measurements for Pb metal with the same applied load to the cell.  In all high-pressure measurements, we recorded the SQUID response curves from the cell with and without the sample and the difference was analyzed to obtain the magnetization from the sample. Details of this procedure are explained in Appendix~\ref{MT4GPa}.

The $^{31}$P NMR measurements were conducted at a fixed field of 5.0 T on three pieces of single crystals ($\sharp$1-3) with the size $\sim$1.0 $\times$ 0.8 $\times$ 0.1 mm$^3$ selected from the same batch. The crystal $\sharp$1, which is the same crystal used in the previous NMR measurements reported in Ref.~\onlinecite{Ishikawa}, was used for the measurements at ambient pressure.  NMR measurements under high pressure at $P$ = 2.6, 4.0 and 6.4~GPa were conducted on the crystal $\sharp$2, while the crystal $\sharp$3 was used for measurements at other pressure values.  An opposed-anvil pressure cell~\cite{Kitagawa} was used for all NMR measurements with glycerol as the pressure medium. The pressure values were determined $in$ $situ$ from the $^{119}$Sn NMR Knight shift measured on Sn metal powder placed in the NMR coil~\cite{Kitagawa}.  

The $^{31}$P NMR spectra were obtained by summing the Fourier transform of spin echo signal recorded at equally spaced frequencies.  The spin lattice relaxation rate $1/T_1$ was measured by saturation recovery method.  The time ($t$) dependence of the spin echo intensity after the saturation pulse was fitted to the following function with a single value of $1/T_1$,
\begin{equation}\label{single}
I(t)=I_{\rm{eq}}\left(1-\exp(-t/T_1)\right), 
\end{equation} 
except for a certain pressure region, where $1/T_1$ has significant distribution as described later.  

Since $^{31}$P nuclei have spin 1/2, each nucleus has a unique magnetic resonance frequency given by 
\begin{eqnarray}{\label{nu_resonance}}
\nu = \gamma_{\rm{N}} |\braket{{\bf{B}}_{\rm{loc}}}| = \gamma_{\rm{N}} |{\bf{B}}_{\rm{ext}}+\braket{{\bf{B}}_{\rm{hf}}}| ,
\end{eqnarray}
where $\gamma_{\rm{N}} = 17.235$ MHz/T is the nuclear gyromagnetic ratio, ${\bf{B}}_{\rm{loc}}$ is the local magnetic field acting on the nucleus, which is the sum of the external field ${\bf{B}}_{\rm{ext}}$ and the microscopic hyperfine field ${\bf{B}}_{\rm{hf}}$ generated by surrounding Mo magnetic moments, and $\braket{\dots}$ denotes the time average. 
In this paper, $^{31}$P NMR spectra are shown as a function of the internal field $B_n$ defined as, 
\begin{equation}{\label{Bn_freq}}
B_n=|{\bf{B}}_{\rm{ext}}+\braket{{\bf{B}}_{\rm{hf}}}|-B_{\rm{ext}}.
\end{equation}
In the paramagnetic state where $\braket{{\bf{B}}_{\rm{hf}}}=0$ at ${\bf{B}}_{\rm{ext}}=0$, 
$B_n$ is proportional to $B_{\rm{ext}}$. Then the NMR shift $K$ is defined as 
\begin{equation}{\label{Kshift}}
K = \frac{B_n}{B_{\rm{ext}}}
\end{equation}

\section{Results}
\subsection {Magnetization}

\begin{figure}[t]
\includegraphics[width=8cm]{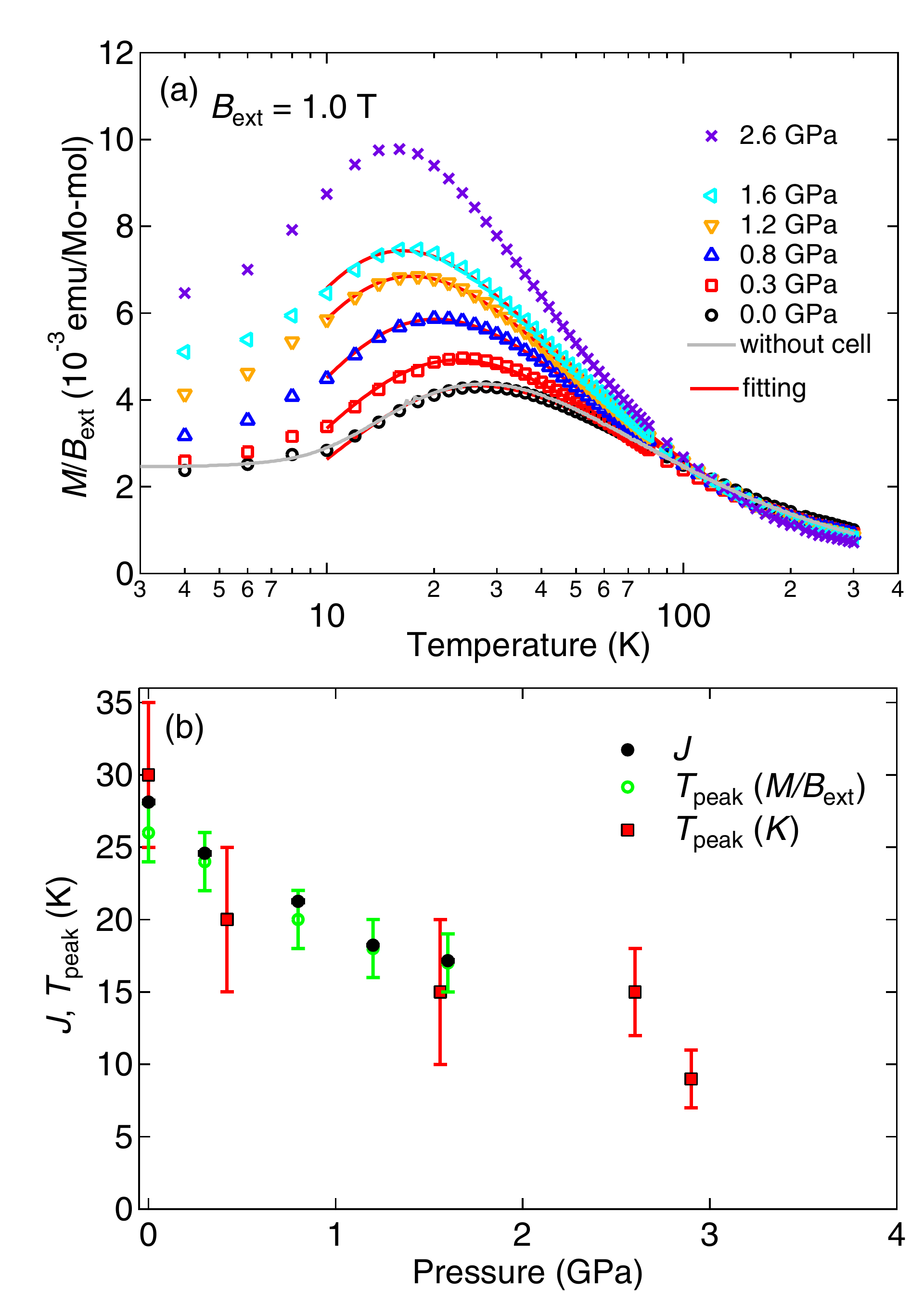}
\caption{\label{fig:MT_piston} (Color online) (a) Temperature dependences of magnetic susceptibility $M/B_{\rm{ext}}$ at $B_{\rm{ext}}$ = 1.0~T for $P \leq 1.6$~GPa measured with a piston-cylinder cell.  The results at 2.6~GPa are obtained by linearly extrapolating the data below 1.6~GPa at each temperature (see Fig.~\ref{fig:MH_extrapolate}).  The red solid curves show fitting to the high-temperature series expansion of the spin-1/2 Heisenberg model on a square lattice with a single exchange parameter $J$ between NN sites.  (b) Pressure dependence of $J$ obtained from the fitting shown in (a) (solid dots).  For comparison, the peak temperatures of $M/B_{\rm{ext}}$ (open circles) and NMR shift $K_c$ (solid squares) are also plotted.} 
\end{figure}

\begin{figure}[t]
\includegraphics[width=8cm]{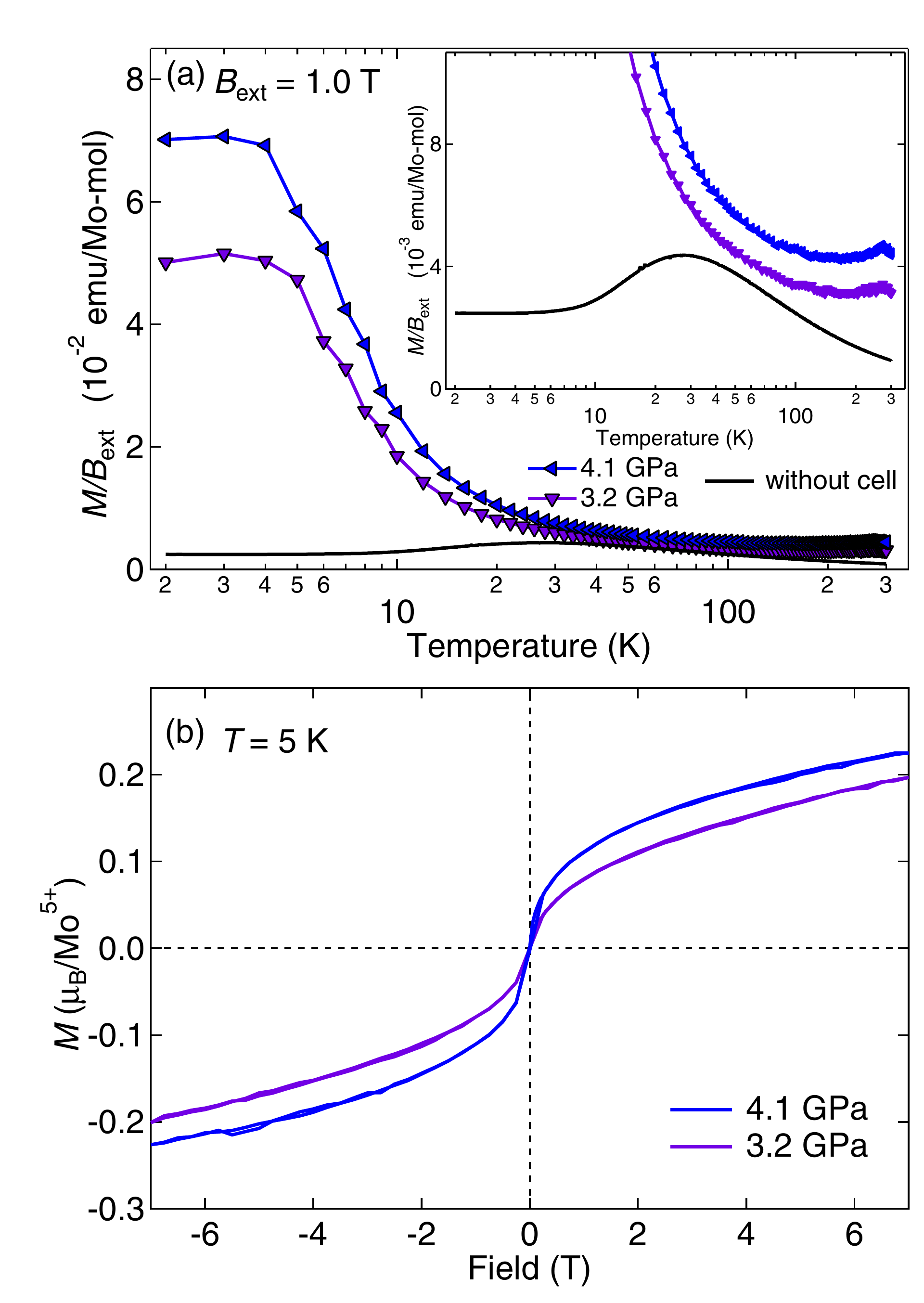}
\caption{\label{fig:MT_tateiwa} (Color online) (a) Temperature dependences of magnetic susceptibility $M/B_{\rm{ext}}$ at $B_{\rm{ext}}$ = 1.0~T for $P \geq 2$~GPa measured with an opposed anvil cell. For comparison $M/B_{\rm{ext}}$ at ambient pressure measured without the pressure cell is also shown.  (b) Magnetization is plotted as a function of field at $T$ = 5~K for $P$ = 3.2 and 4.1~GPa.} 
\end{figure}

\begin{figure}[t]
\includegraphics[width=8cm]{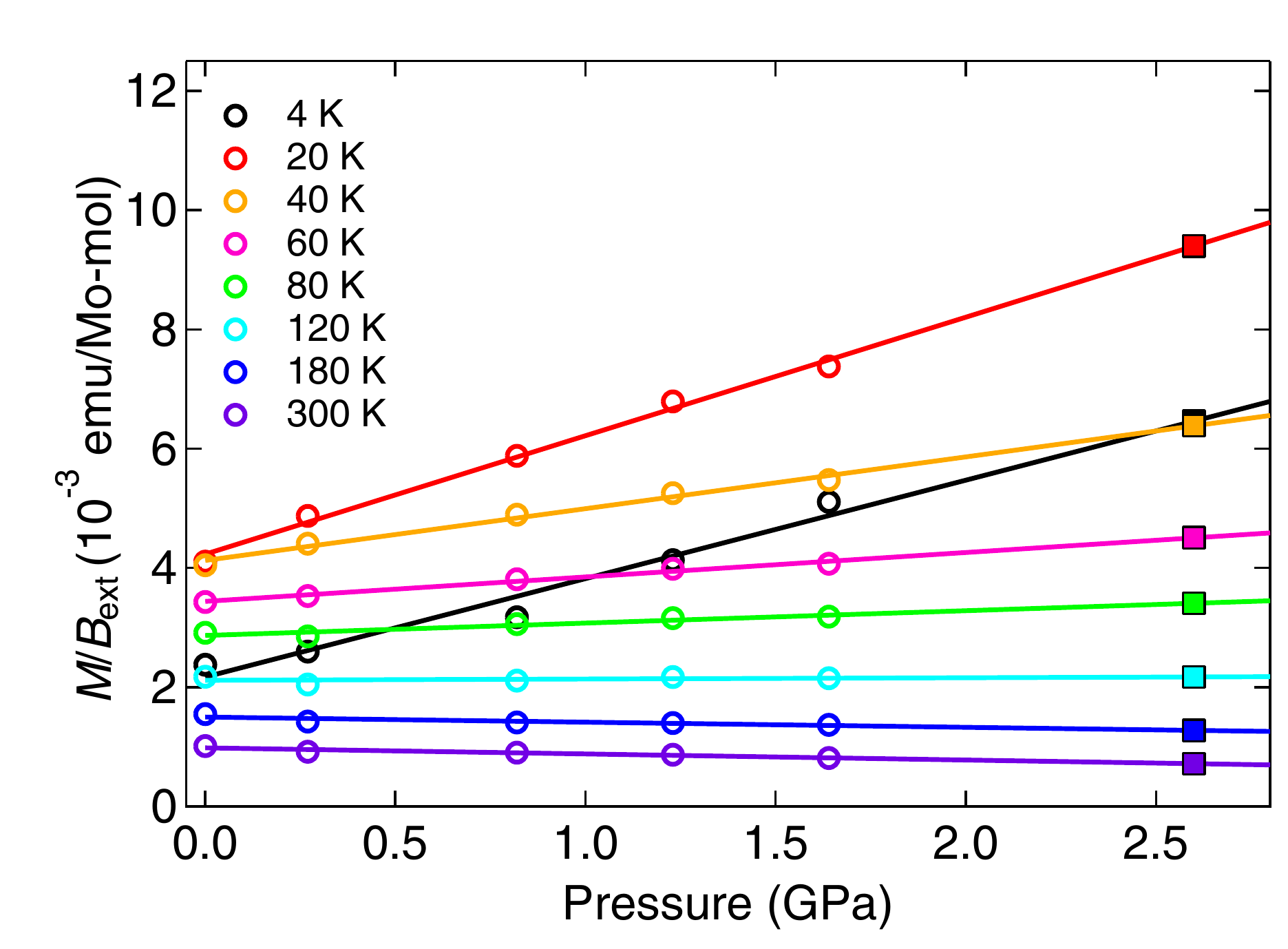}
\caption{\label{fig:MH_extrapolate} (Color online) Pressure dependences of $M/B_{\rm{ext}}$ at selected temperatures (open circles) fitted to straight lines. The solid squares show the extrapolated values of $M/B_{\rm{ext}}$ at 2.6 GPa.}.
\end{figure}

Figure \ref{fig:MT_piston}(a) shows the temperature dependences of magnetic susceptibility $M/B_{\rm{ext}}$ at $B_{\rm{ext}}$ = 1.0~T for $P = 0.0 \sim 1.6$~GPa measured with a piston-cylinder pressure cell.  With increasing pressure, $M/B_{\rm{ext}}$ increases and the broad peak observed near 30 K at ambient pressure shifts to lower temperatures. The peak temperature of $M/B_{\rm{ext}}$ is plotted against pressure in Fig.~\ref{fig:MT_piston}(b).  The exchange interaction at each pressure was determined by fitting $M/B_{\rm{ext}}$ to the high-temperature series expansion of the spin-1/2 Heisenberg model on a square lattice with a single exchange parameter $J$~\cite{Roshbrooke} as described in the previous report~\cite{Ishikawa}. The fitting is satisfactory down to 10~K for all pressure values [see the lines in Fig.~\ref{fig:MT_piston}(a)], indicating that $J_2$ remains dominant over $J_1$ up to 1.6 GPa as is the case at ambient pressure. The pressure dependence of $J$ plotted in Fig.~\ref{fig:MT_piston}(b) agrees very well with the peak temperature of $M/B_{\rm{ext}}$, indicating that the latter provides a good quantitative estimate of $J_2$ in this material.  

Figure \ref{fig:MT_tateiwa}(a) shows $M/B_{\rm{ext}}$ at higher pressures measured with an opposed-anvil cell. Details of the analysis of the raw data of SQUID response at 4.1~GPa is described in Appendix {\ref{MT4GPa}}. For comparison, $M/B_{\rm{ext}}$ at ambient pressure measured without the pressure cell is also shown. Because of limited sample volume and large background from the cell with inevitable uncertainty, it is likely that the results at high temperatures are influenced by systematic error. However, the dramatic increase of $M/B_{\rm{ext}}$ at high pressures and low temperatures is beyond such uncertainty.  
At $P$ = 3.2 and 4.1~GPa, $M/B_{\rm{ext}}$ increases rapidly below 10~K. Also the manetization curves at $T$ = 5~K shown in Fig.~\ref{fig:MT_tateiwa}(b) exhibits abrupt rise near $B_{\rm{ext}} \sim 0$, followed by nonlinear and gradual increase at higher fields. Such behaviors are typical of a ferromagnet. Since the magnitude of the initial rise of magnetization is less than 0.1~$\mu_{\rm{B}}$, this weak ferromagnetism is most likely caused by canting of antiferromagnetic moments.    

Since it is difficult to obtain reliable data for $1.6 \leq P \leq 3.2$~GPa, we have attempted to extrapolate the results below 1.6~GPa obtained with the piston-cylinder cell to higher pressures. In Fig.~\ref{fig:MH_extrapolate}, the pressure dependences of $M/B_{\rm{ext}}$ are shown at selected temperature values (open circles). The data can be well fitted to straight lines at all temperatures. This allows us to estimate $M/B_{\rm{ext}}$ at higher pressures by extrapolating the linear relations. The plot of $M/B_{\rm{ext}}$ at 2.6~GPa in Fig.~\ref{fig:MT_piston}(a) is an example of such extrapolation, which will be used in the analysis of NMR shift data in Sec.~\ref{shift} 

\subsection{$^{31}$P NMR spectra}\label{spectra}
\begin{figure*}[t]
\includegraphics[width=17cm]{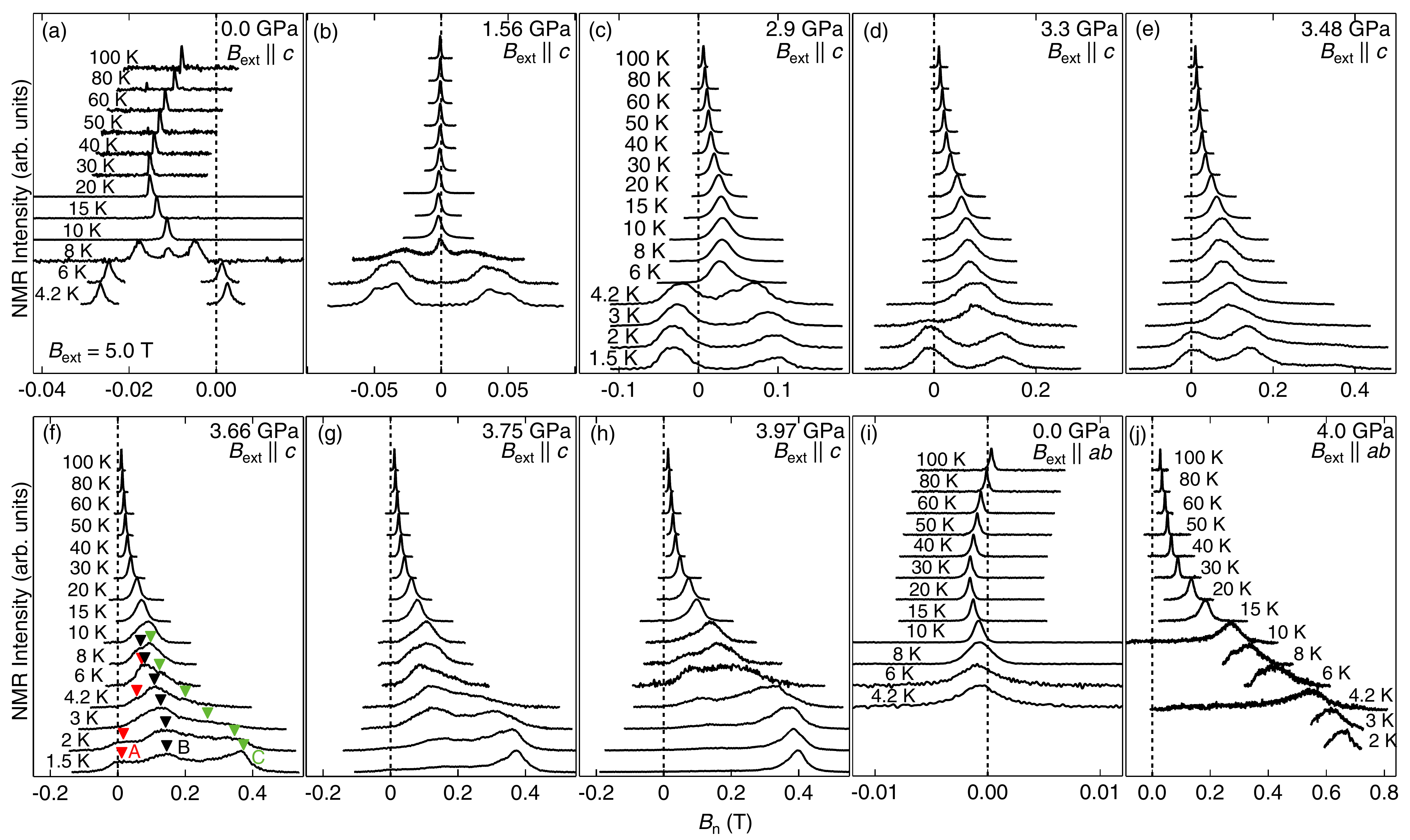}
\caption{\label{fig:spc_pressure_Dependence} (Color online) Temperature dependences of $^{31}$P NMR spectra obtained for $B_{\rm{ext}}$ = 5.0 T along the $c$-axis (a-h) and in the $ab$-plane (i-j) under various pressures. The red, black, and blue arrows in (f) represent the three peaks (A, B and C) of the spectra, at which $1/T_1$ was measured.}
\end{figure*}

Figures~\ref{fig:spc_pressure_Dependence}(a)$-$(h) show the temperature dependences of $^{31}$P NMR spectra obtained for $B_{\rm{ext}}$ = 5.0 T along the $c$-axis under various pressures. The spectra are plotted as a function of the internal field $B_n$ defined in Eq.~(\ref{Bn_freq}). At ambient pressure [Fig.~\ref{fig:spc_pressure_Dependence}(a)], the resonance peak appears at a negative value of $B_n$ in the paramagnetic phase down to 10~K. This means that the direction of the hyperfine field at $^{31}$P nuclei produced by the surrounding Mo moments is opposite to the external field. The temperature dependence of $|B_n|$ at the spectral peak is similar to that of the magnetic susceptibility with a maximum near 30 K.  Below $T_{\rm{N}}$ = 8 K the spectrum splits into two peaks as has been reported in Ref.~[\onlinecite{Ishikawa}]. This splitting is ascribed to the CAF type magnetic order (see Ref.~[\onlinecite{Ishikawa}] and the discussion in Sec.~\ref{Magnetic structure}). At $P$ = 1.56~GPa [Fig.~\ref{fig:spc_pressure_Dependence}(b)], the peak value of $B_n$ stays nearly zero in the entire temperature range of the paramagnetic phase. With further increase of pressure, the sign of $B_n$ is reversed to positive as indicated by the spectra (c)-(h). The line splitting at low temperatures continues up to 3.3 GPa. However, the onset temperature of the splitting decreases gradually with pressure above 2~GPa.  

\begin{figure}[t]
\includegraphics[width=8cm]{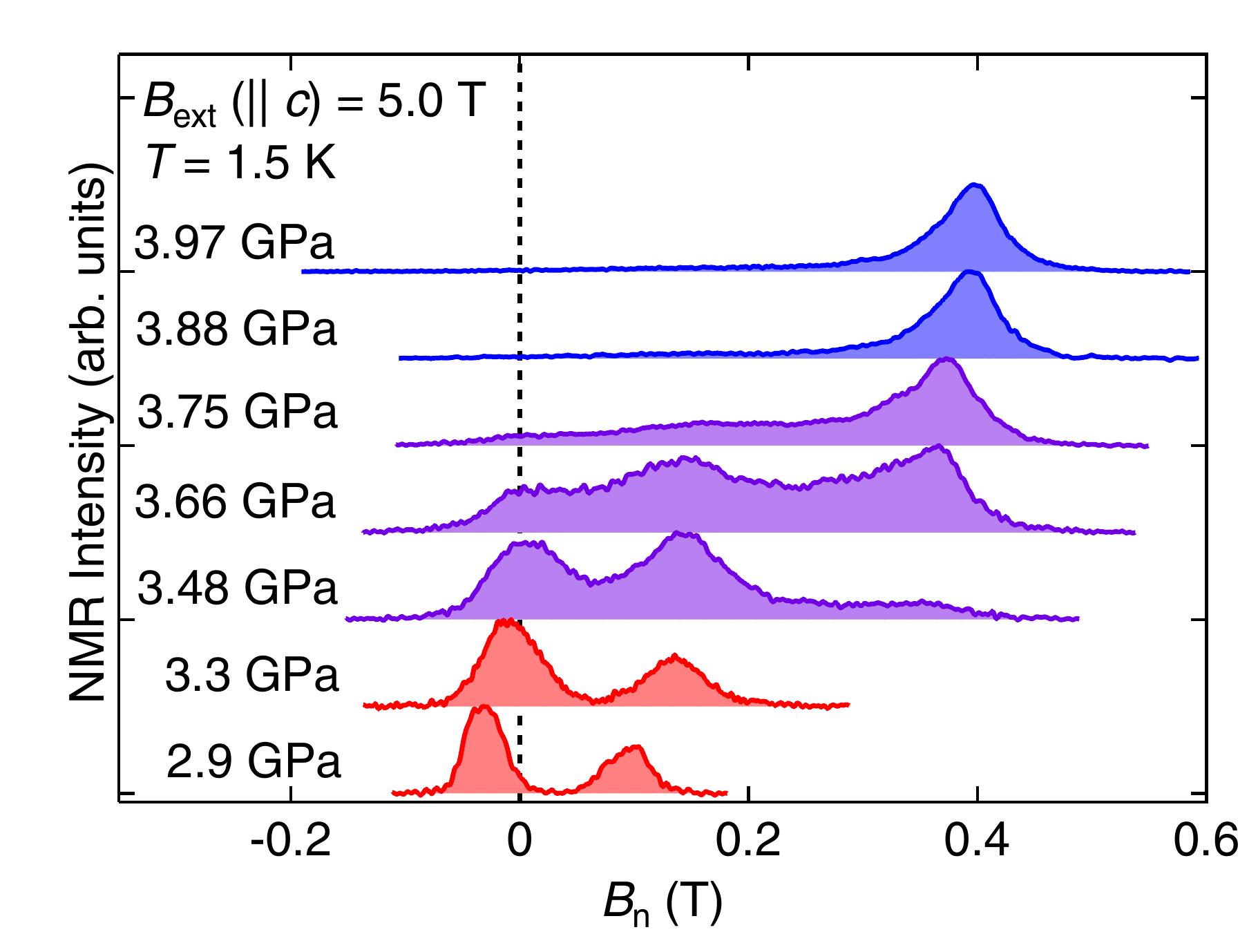}
\caption{\label{fig:spc_1_5K_P_dependence} (Color online) $^{31}$P NMR spectra obtained at 1.5 K in the magnetic field of 5.0 T along the $c$ axis under various pressures between 2.9 and 3.97~GPa.} 
\end{figure}

The spectral shape in the ordered phase changes remarkably above 3.3~GPa. To make it easier to see this change, the spectra at the lowest temperature (1.5~K) are collected in Fig.~\ref{fig:spc_1_5K_P_dependence} for $2.9 \leq P \leq 4$~GPa.  Up to 3.3~GPa, the spectra consist of split two peaks originating from the CAF order. For $P$ = 3.48~GPa, a new small peak appears at $B_n$ = 0.37 T [Figs.~\ref{fig:spc_pressure_Dependence}(e) and \ref{fig:spc_1_5K_P_dependence}]. With increasing pressure, this new peak grows while the other two peaks from the CAF phase get suppressed, until the spectrum has only one peak near $B_n$ = 0.4~T above 3.88~GPa. As we discuss later in Sec.~\ref{Magnetic structure}, this new peak is assigned to the NAF order.  The spectra in the intermediate pressure region between 3.3 and 3.9 GPa can be understood as the superposition of two spectra, each of which belongs to CAF and NAF phase, although the individual lines are significantly broader than those in the spectra of pure phases. See, for example, the spectrum at 3.66~GPa. This suggests that two phases coexist in a finite range of pressure. We will come back to this point later. 

\begin{figure}[t]
\includegraphics[width=8.5cm]{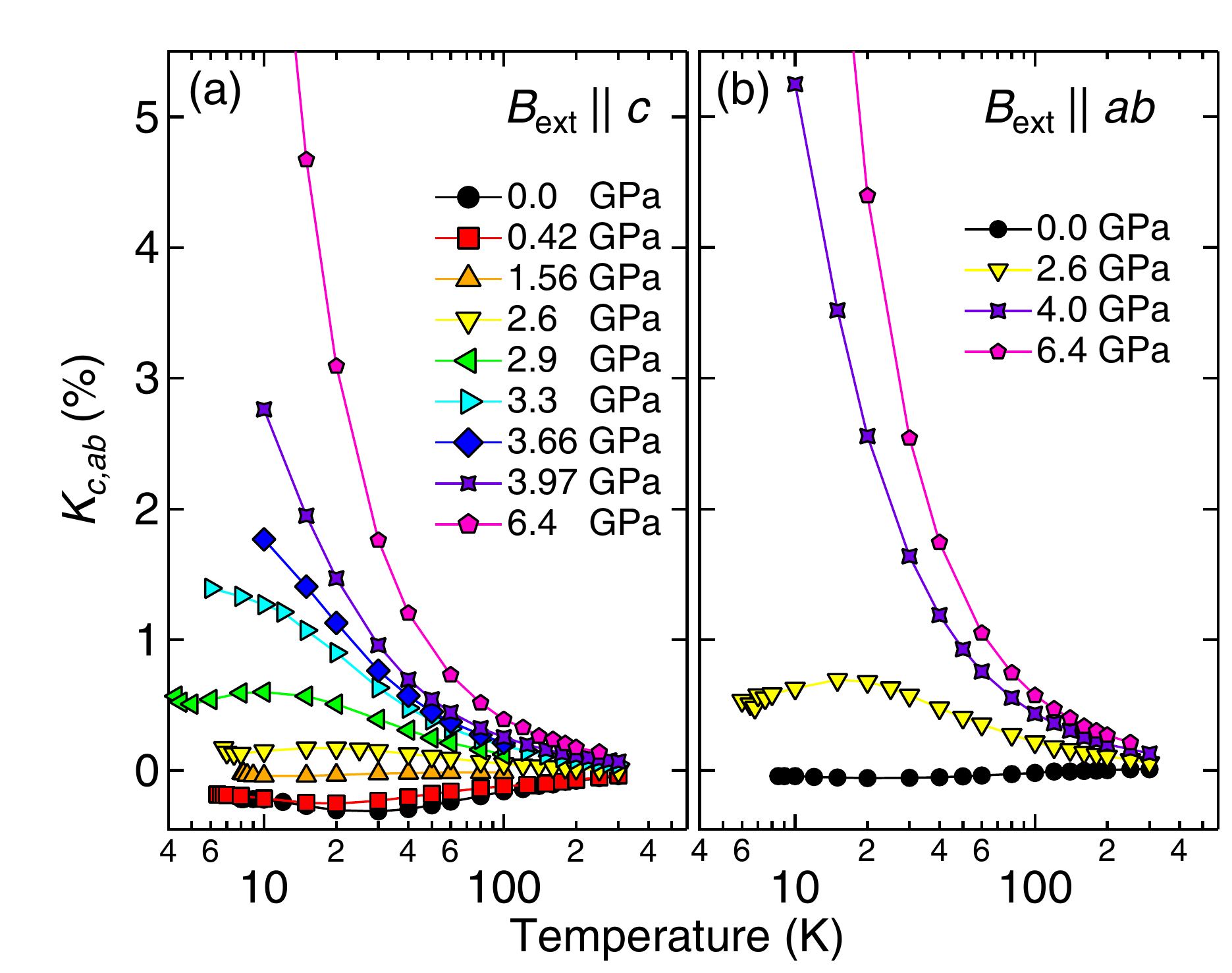}
\caption{\label{fig:K_pressure} (Color online) Temperature dependences of NMR shift $K$ measured under various pressures in the magnetic field of 5.0 T (a) along the $c$ axis, and (b) in the $ab$ plane.} 
\end{figure}

\begin{figure}[t]
\includegraphics[width=8cm]{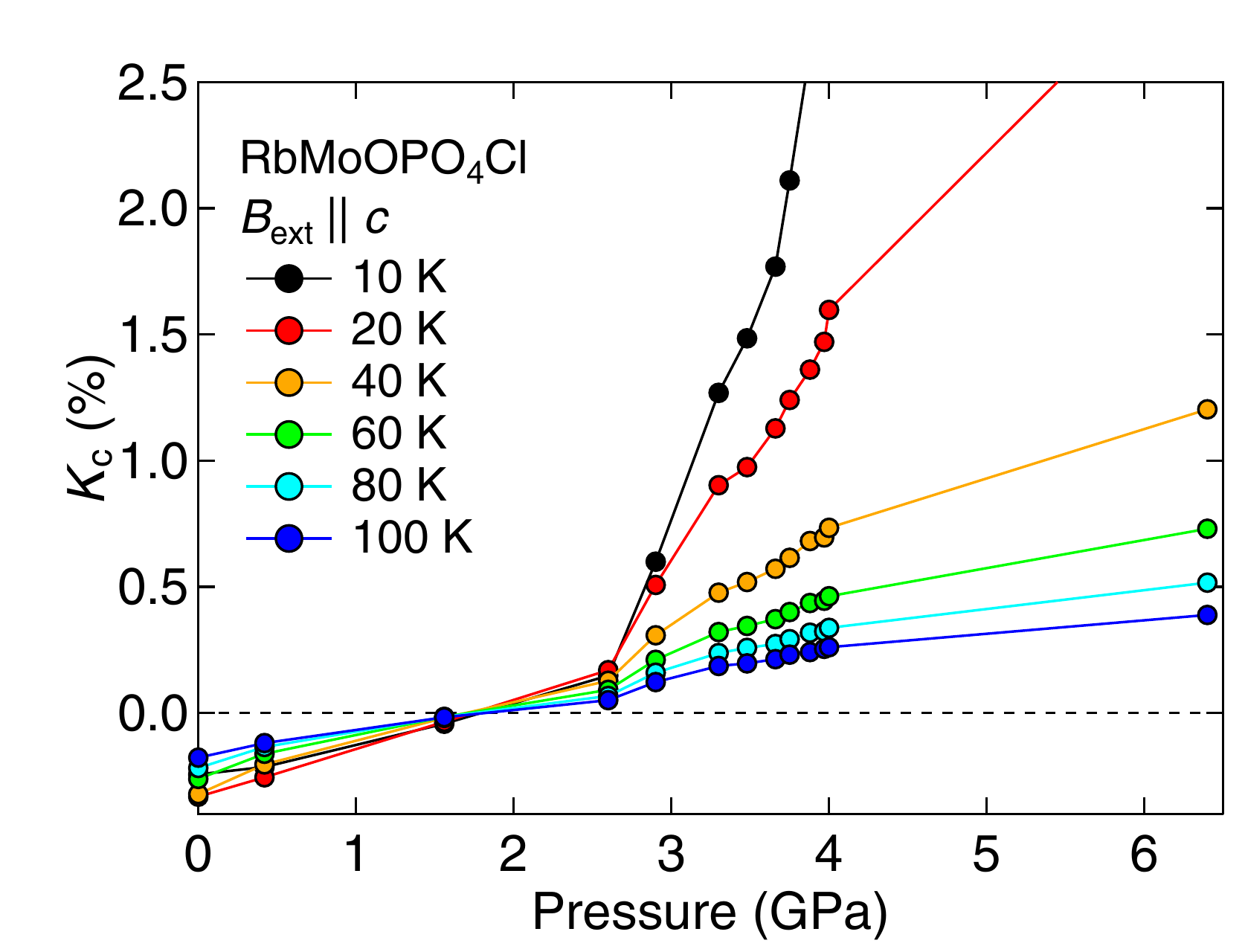}
\caption{\label{fig:K_P} (Color online) Isothermal pressure dependences of $K_c$ at various temperatures. The solid lines are guides for the eyes.} 
\end{figure}

The temperature dependence of NMR spectra obtained for $B_{\rm{ext}}$ = 5.0 T parallel to the $ab$-plane are shown in Fig.~\ref{fig:spc_pressure_Dependence}(i) at ambient pressure and (j) at 4.0~GPa.  At ambient pressure, the peak of the spectrum remains near $B_n$ = 0 without splitting below $T_{\rm{N}}$, in contrast to the case for ${\bf{B}}_{\rm{ext}}\parallel c$. This has been explained by the CAF order with the antiferromagnetic moments in the $ab$-plane~\cite{Ishikawa}.  The spectrum at 4.0 GPa does not show any splitting either.  However, the spectral peak at low temperatures below 4.2 K appears at a large value of $B_n$ exceeding 0.6 T, indicating a large uniform internal field parallel to the external field.  

\subsection{$^{31}$P NMR shift}\label{shift}
From the value of $B_n$ at the peak of the NMR spectra, the NMR shift $K$ is determined by Eq.~(\ref{Kshift}) above $T_{\rm{N}}$. Figure~\ref{fig:K_pressure} shows the temperature dependence of the shift for ${\bf{B}}_{\rm{ext}} \parallel c$ [$K_c$ in (a)] and for ${\bf{B}}_{\rm{ext}} \parallel ab$ [$K_{ab}$ in (b)] under various pressures. At ambient pressure, the shift stays unchanged when ${\bf{B}}_{\rm{ext}}$ is rotated in the $ab$ plane because of the $S_4$ symmetry at P sites in the $P4/nmm$ space group. We have confirmed that this isotropy of $K_{ab}$ in the $ab$ plane is maintained even at the highest pressure (6.4 GPa) of our measurements, as presented in Appendix~\ref{6.4GPa}. This indicates that $S_4$ symmetry at P site is preserved under high pressure. 

The sign of $K_c$ changes from negative to positive at a relatively low pressure near 1.6~GPa, as already noticed from the variation of NMR spectra in Fig.~\ref{fig:spc_pressure_Dependence}. With further increasing pressure, $K_c$ exhibits remarkable enhancement. The broad peak of $|K_c|$ near 30~K observed at ambient pressure shifts to lower temperatures at higher pressures similar to the behavior of susceptibility. The pressure dependence of the peak temperature of $|K_c|$ plotted in Fig.~\ref{fig:MT_piston}(b) indicates that reduction of $J_2$ with pressure continues up to 3~GPa. Above 3.3~GPa, $K_c$ does not show the broad peak any more but monotonically increases down to the lowest temperature. A similar behavior is observed for $K_{ab}$, whose magnitude is larger than $K_c$.   

Quite surprisingly, the same set of data exhibit striking anomaly when plotted as a function of pressure. In Fig.~\ref{fig:K_P} the isothermal pressure dependence of $K_c$ is displayed at various temperatures from 10 to 100 K.  All curves show a clear kink or a sudden increase of slope at a common pressure value of 2.6~GPa. The kink becomes more pronounced at lower temperatures. Such an anomaly strongly suggests a pressure-induced phase transition. Since the anomaly persists up to as high as 100~K, it is likely a structural transition not a magnetic one.  We notice that the critical pressure of 2.6~GPa is substantially smaller than the pressure range of 3.3 $-$ 3.9~GPa, where the change of AF spin structure is inferred from the NMR spectra. This suggests that the structural transition triggers a drastic change of the ratio $J_2/J_1$, which then leads to the change of AF spin structure at higher pressures.    

Let us now relate the measured shift to the microscopic hyperfine coupling tensor between $^{31}$P nuclei and Mo magnetic moments. Generally in magnetic insulators, the internal field at ligand nuclei is produced via the transferred hyperfine interaction mediated by the hybridization between $d$ orbitals of the magnetic ions and $p$ or $s$ orbitals of ligand ions. Since such an interaction is short ranged, it is sufficient to consider only the contribution from nearest neighbors. Then the hyperfine field ${\bf{B}}_{\rm{hf}}$ at a $^{31}$P nucleus in RbMoOPO$_4$Cl, can be expressed as the sum of contributions over four nearest-neighbor Mo moments~\cite{Ishikawa},  
\begin{equation}{\label{hfcoupling}}
{\bf{B}}_{\rm{hf}}= \sum_{i=1}^{4}{\bf{A}}^{\alpha}({\bf{r}}_i) \cdot {\bm{\mu}}({\bf{r}}_i),
\end{equation}
where ${\bf{A}}^{\alpha}({\bf{r}}_i)$ is the transferred hyperfine coupling tensors between the P nucleus at $\alpha$-site and the Mo magnetic moment ${\bm{\mu}}({\bf{r}}_i)$ (with $\mu_{\rm{B}}$ as a unit) at ${\bf{r}}_i$.
We write the hyperfine coupling tensor between P(A) and Mo(1) sites in Fig.~\ref{fig:RbMoOPO4Cl_MoOPO4}(b) as  
\begin{eqnarray}
{\bf{A}}^{\rm{A}}({\bf{r}}_1)&=\left(\begin{array}{ccc}
A_{aa} & A_{ab} & A_{ac}\\
A_{ba} & A_{bb} & A_{bc}\\
A_{ca} & A_{cb} & A_{cc}
\end{array}\right) {\label{A1}}.
\end{eqnarray}
We should remark that among nine components of ${\bf{A}}^{\rm{A}}({\bf{r}}_1)$ tensor, four of them, $A_{ab}$, $A_{ba}$, $A_{bc}$, and $A_{cb}$, should be zero in the $P4/nmm$ crystal structure of RbMoOPO$_4$Cl at ambient pressure, since the bond connecting P(A) and Mo(1) sites are on the mirror plane perpendicular to the $b$-axis. However, we keep these components in Eq.~(\ref{A1}) because the crystal structure at high pressures may break the mirror symmetry as will be discussed later.   

 The coupling tensors ${\bf{A}}^{\rm{A}}({\bf{r}}_i)$ ($i = 2-4$) for the other three bonds, P(A)-Mo($i$) ($i = 2-4$), can be obtained by transforming ${\bf{A}}^{\rm{A}}({\bf{r}}_1)$ successively with $S_4$ operation,   
\begin{eqnarray}
{{\bf{A}}}^{\rm{A}} ({\bf{r}}_2)&= \left(\begin{array}{ccc}
A_{bb} & -A_{ba} & A_{bc}\\
-A_{ab} & A_{aa} & -A_{ac}\\
A_{cb} & -A_{ca} & A_{cc}
\end{array}\right){\label{A2}}, \\
{{\bf{A}}}^{\rm{A}} ({\bf{r}}_3)&= \left(\begin{array}{ccc}
A_{aa} & A_{ab} & -A_{ac}\\
A_{ba} & A_{bb} & -A_{bc}\\
-A_{ca} & -A_{cb} & A_{cc}
\end{array}\right){\label{A3}}, \\
{{\bf{A}}}^{\rm{A}} ({\bf{r}}_4)&= \left(\begin{array}{ccc}
A_{bb} & -A_{ba} & -A_{bc}\\
-A_{ab} & A_{aa} & A_{ac}\\
-A_{cb} & A_{ca} & A_{cc}
\end{array}\right){\label{A4}}.
\end{eqnarray}

In the paramagnetic phase, the thermal average of the moments are uniform and given by $\braket{{\bm{\mu}}({\bf{r}}_i)} = {\bm{\chi}} \cdot {\bf{B}}_{\rm{ext}} / N_{\rm{A}}\mu_{\rm{B}}$ for all $i$, where ${\bm{\chi}}$ is the magnetic susceptibility tensor per mole and $N_{\rm{A}}$ is Avogadro number. The hyperfine field can then be expressed as 
\begin{equation}{\label{tr_hyperfine}}
\braket{{\bf{B}}_{\rm{hf}}}= \frac{\left( \sum_{i=1}^{4}{\bf{A}}^{\rm{A}}({\bf{r}}_i) \right) \cdot {\bm{\chi}} \cdot {\bf{B}}_{\rm{ext}}}{N_{\rm{A}}\mu_{\rm{B}}}= {\bf{K}}\cdot{\bf{B}}_{\rm{ext}},
\end{equation}
where the shift tensor ${\bf{K}}$ is given by  
\begin{equation}{\label{K_A}}
{\bf{K}}=\left(\begin{array}{ccc}
2(A_{aa}+A_{bb}) & 2(A_{ab}-A_{ba}) & 0\\
-2(A_{ab}-A_{ba}) & 2(A_{aa}+A_{bb}) & 0\\
0 & 0 & 4A_{cc}
\end{array}\right) \cdot \frac{{\bm{\chi}}}{N_{\rm{A}}\mu_{\rm{B}}}.
\end{equation} 
The off-diagonal component $2(A_{ab}-A_{ba})$ should be much smaller than the diagonal component $2(A_{aa}+A_{bb})$, even if it is nonzero, and produces a hyperfine field perpendicular to the external field. From Eqs.~(\ref{Bn_freq}) and (\ref{Kshift}), this makes only a second order contribution to $B_n$ and $K$. Therefore we can neglect the off-diagonal term.     
We then obtain 
\begin{equation}{\label{K_c_ab}}
K_c = \frac{4A_{cc}\chi_{\parallel}}{N_{\rm{A}}\mu_{\rm{B}}}, \ \ K_{ab} = \frac{2(A_{aa}+A_{bb})\chi_{\perp}}{N_{\rm{A}}\mu_{\rm{B}}},
\end{equation}
where $\chi_{\parallel}$ and $\chi_{\perp}$ are the magnetic susceptibility for ${\bf{B}}_{\rm{ext}}\parallel c$ and ${\bf{B}}_{\rm{ext}}\parallel ab$, respectively.
Thus both $K_{ab}$ and $K_{c}$ are expected to be proportional to the magnetic susceptibility and the hyperfine coupling constants $A_{aa}+A_{bb}$ and $A_{cc}$ can be determined from the proportionality constants.  Since the susceptibility was measured for powder sample, uniaxial anisotropy of $\chi$ due to anisotropic $g$ factor have to be absorbed in the anisotropy of the hyperfine coupling constants.  

\begin{figure}[t]
\includegraphics[width=8cm]{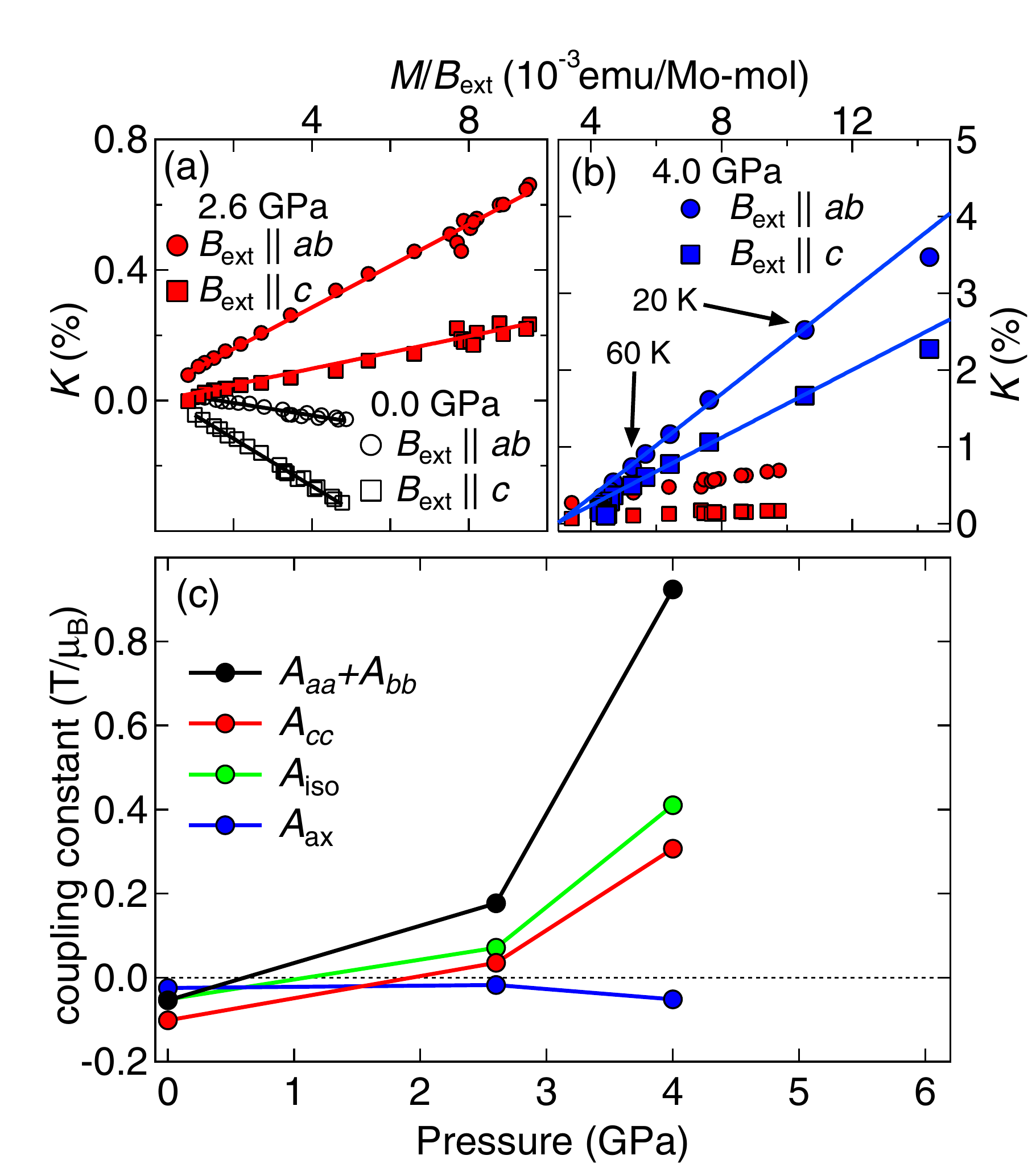}
\caption{\label{fig:K_chi_plot} (Color online) [(a) and (b)] The shifts $K_c$ and $K_{ab}$ are plotted against $M/B_{\rm{ext}}$ with temperature as an implicit parameter for the pressure values of 0.0, 2.6, and 4.0~GPa. The data are fitted to straight lines. The fitting range for the 4.0~GPa data is chosen between 20 and 60~K.  (c) Pressure dependences of the hyperfine coupling constants $A_{aa}+A_{bb}$, $A_{cc}$, $A_{\rm{iso}}$, and $A_{\rm{ax}}$.} 
\end{figure}

The shifts $K_c$ and $K_{ab}$ are plotted against the susceptibility with temperature as an implicit parameter in Fig.~\ref{fig:K_chi_plot} for $P$ = 0.0, 2.6, and 4.0~GPa.  In these plots, the values of $B_{\rm{ext}}$ in Eqs.~(\ref{Bn_freq}) and (\ref{Kshift}) are corrected for the Lorentz and demagnetization fields produced by the bulk magnetization based on the geometry of the crystals and the magnetization data. All sets of data can be fitted to straight lines and the coupling constants $A_{aa}+A_{bb}$ and $A_{cc}$ are determined as listed in Table~\ref{tab:table1} and plotted in Fig.~\ref{fig:K_chi_plot}(c). The good quality of fitting should justify the extrpolation procedure to estimate the susceptibility at 2.6~GPa [Figs.~\ref{fig:MT_piston}(a) and \ref{fig:MH_extrapolate}].
 
\begin{table}[b]
\caption{\label{tab:table1} The hyperfine coupling constants in unit of T/$\mu_{\rm{B}}$ for the pressure values of 0.0 (ambient pressure), 2.6, and 4.0~GPa.  The isotropic and axial anisotropic part of the hyperfine coupling constants, $A_{\rm{iso}}$ and $A_{\rm{ax}}$ are defined by Eq.~(\ref{A_iso,ax}).} 
\begin{ruledtabular}
\begin{tabular}{ccccc}
$P$ (GPa) & $A_{aa}+A_{bb}$ & $A_{cc}$ & $A_{\rm{iso}}$ & $A_{\rm{ax}}$ \\
\hline
0.0 & $-$0.054 &$-$0.102 & $-$0.052 & $-0.025$  \\
2.6 & 0.177 & 0.035 & 0.071 & $-$0.018 \\
4.0 & 0.924 & 0.307 & 0.410 & $-$0.052 \\
\end{tabular}
\end{ruledtabular}
\end{table}

Both $A_{aa}+A_{bb}$ and $A_{cc}$ increase drastically above the critical pressure of 2.6~GPa, indicating that the pressure-induced phase transition causes substantial change in the hybridization between Mo-$d$ and P-$s$, $p$ orbitals. To obtain orbital-selective information, we define the isotropic and axially anisotropic components of the hyperfine coupling tensor,  $A_{\rm{iso}}$ and $A_{\rm{ax}}$, by  
\begin{eqnarray}\label{A_iso,ax}
A_{\rm{iso}}=\left(A_{aa}+A_{bb}+A_{cc}\right)/3, \nonumber \\
A_{\rm{ax}}=\left(2A_{cc}-(A_{aa}+A_{bb})\right)/6.
\end{eqnarray}
Generally, the spin density on $s$ orbitals produces positive isotropic hyperfine coupling.
In contrast, the spin density on $p$ orbitals generates both anisotropic coupling via spin dipolar field and negative isotropic coupling due to core polarization effect\cite{abragam_b}. As shown in Fig.~\ref{fig:K_chi_plot}(c), $A_{\rm{iso}}$ changes sign from negative to positive near 1~GPa and becomes strongly enhanced above 2.6 GPa, while $A_{\rm{ax}}$ changes only modestly with pressure. This result indicates that the structural transition at 2.6~GPa promotes spin transfer to the $s$ orbitals at P sites. 

\begin{figure}[t]
\includegraphics[width=8.5cm]{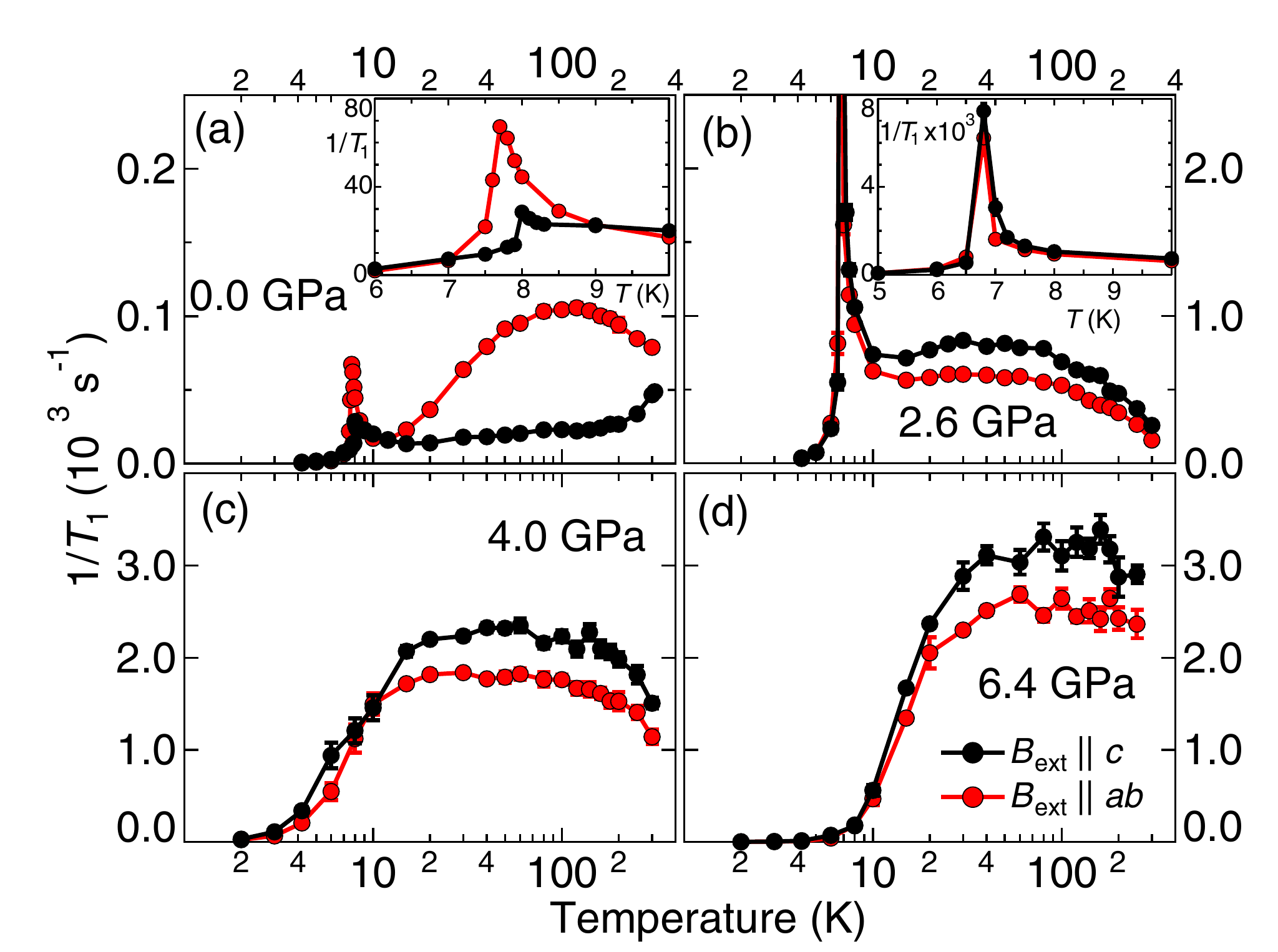}
\caption{\label{fig:T1_aniso} (Color online) Temperature dependences of $1/T_1$ at $B_{\rm{ext}}$ = 5.0 T parallel to the $c$-axis (black circles) or in the $ab$-plane (red circles) under various pressures.} 
\end{figure}

\begin{figure}[t]
\includegraphics[width=8.5cm]{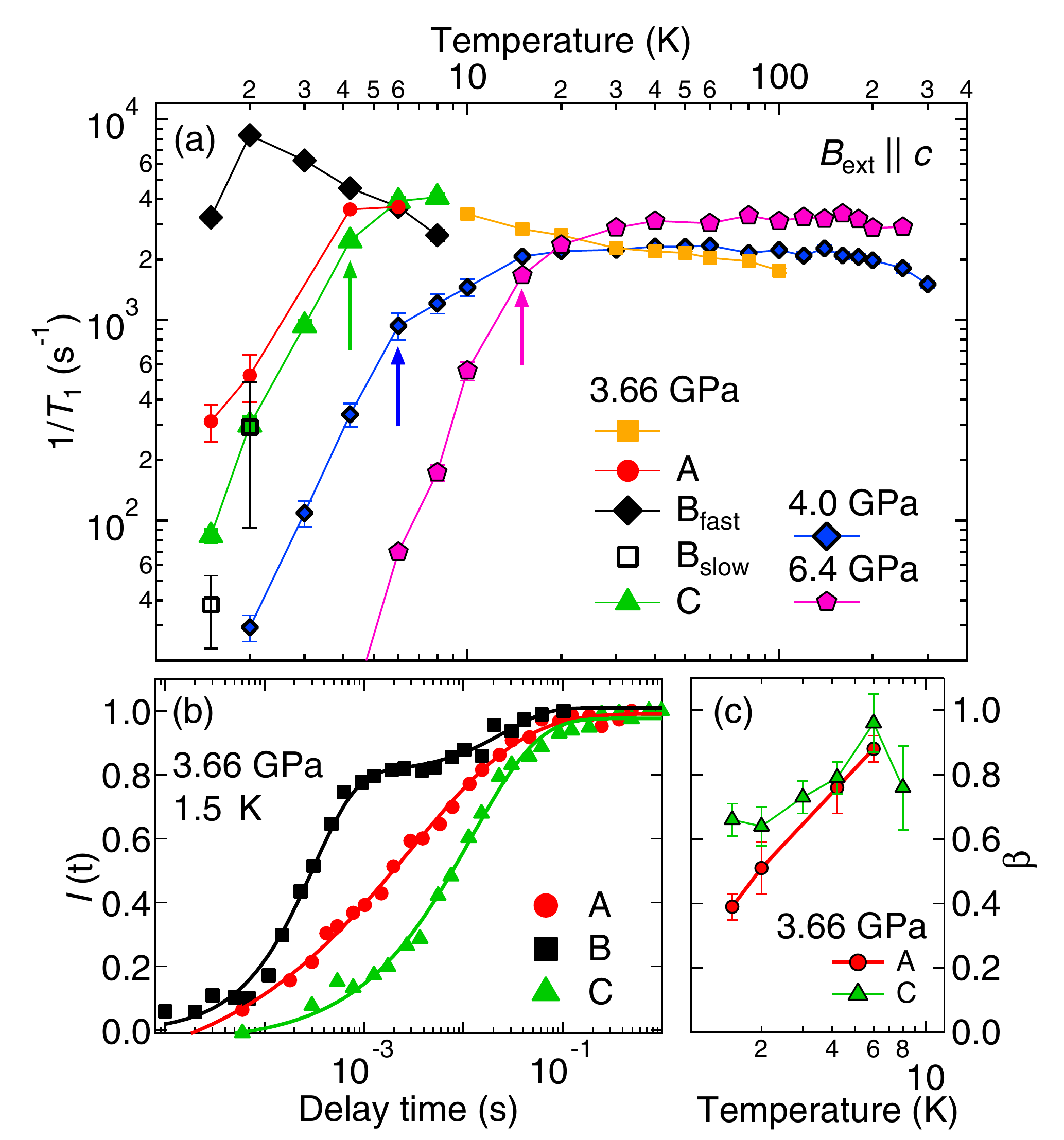}
\caption{\label{fig:T1_intermediate2} (Color online) (a) Temperature dependence of $1/T_1$ at $B_{\rm{ext}}$ = 5.0 T parallel to the $c$-axis for the pressure values $P$ = 3.66, 4.0, and 6.4~GPa.  At $P$ = 3.66~GPa and low temperatures below 10~K, $1/T_1$ was measured at the three peaks of the spectrum marked as A, B and C in Fig.~\ref{fig:spc_pressure_Dependence}(f). The onset temperatures of rapid decrease of $1/T_1$ indicated by the arrows are taken as the magnetic transition temperature of the NAF order. (b) The recovery curves of the nuclear magnetization at 1.5 K for $P$ = 3.66~GPa. (c) Temperature dependence of the stretch exponent $\beta$ for the peak A and C.} 
\end{figure}

\subsection{Spin lattice relaxation rate}\label{spin lattice relaxation rate}
Figure~\ref{fig:T1_aniso} shows the temperature dependences of nuclear spin lattice relaxation rate measured at $B_{\rm{ext}}$ = 5.0 T along the $c$-axis, $(1/T_1)_c$, and in the $ab$-plane, $(1/T_1)_{ab}$, under various pressures. Let us first remark on the very unusual anisotropic behavior at ambient pressure. Since $J_1$ is nearly zero at ambient pressure, we expect the spin system to be decoupled into two independent square lattices, each of which has the unfrustrated NN interaction $J=$ 29~K. Then $1/T_1$ should be independent of temperature for $T \gg J$\cite{Moriya1, Moriya2, Singh1, Gelfand}.    

On the contrary, the $1/T_1$ data in Fig.~~\ref{fig:T1_aniso}(a) exhibit strong temperature dependence even near the room temperature ($T \sim$~10$J$). Furthermore, the temperature dependence is remarkably anisotropic. While $(1/T_1)_{ab}$ exhibits a broad peak near 100~K, $(1/T_1)_c$  shows rapid decrease from 300~K down to 200~K, followed by mild decrease at lower temperatures. We discuss these results in more detail in Appendix~\ref{T1_ambient} based on a phenomenological model. At low temperatures, $(1/T_1)_{ab}$ shows a sharp peak at $T_{\rm{N}}$ = 8~K, which should be ascribed to the divergence of antiferromagnetic correlation length and associated critical slowing down of the spin fluctuations.  Although $(1/T_1)_c$ also shows a peak at $T_{\rm{N}}$, the peak value of $(1/T_1)_c$ is much smaller than that of $(1/T_1)_{ab}$. 

With increasing pressure up to 2.6 GPa, both $(1/T_1)_{ab}$ and $(1/T_1)_c$ are strongly enhanced  and become more isotropic [Fig.~~\ref{fig:T1_aniso}(b)]. They also show strong peaks at $T_{\rm{N}}$ = 6.8~K, which are much more pronounced than at ambient pressure and nearly isotropic.   

Further increase of pressure causes drastic change of behavior.  At 4.0 and 6.4~GPa, the sharp peak at $T_{\rm{N}}$ is no more observed either in $(1/T_1)_{ab}$ or in $(1/T_1)_c$ but both decrease steeply below about 20~K.  We should emphasize that this change of behavior of $1/T_1$ occurs synchronously with the change of NMR spectrum in the ordered states as shown in Fig.~\ref{fig:spc_1_5K_P_dependence}. In fact, we demonstrate in Sec.~\ref{T1_discussion} that the presence (absence) of the divergent peak of $1/T_1$ can be understood by the dominant CAF (NAF) type spin fluctuations. 

In the intermediate pressure range between 3.4 and 3.8~GPa, the NMR spectra consist of multiple peaks (Fig.~\ref{fig:spc_1_5K_P_dependence}). Therefore, we measured $1/T_1$ at each peak separately. Figure \ref{fig:T1_intermediate2} (a) shows the temperature dependences of $(1/T_1)_c$ at 3.66~GPa measured at the three peaks marked as A, B, and C in Fig.~\ref{fig:spc_pressure_Dependence}(f). At this pressure and temperatures below 4~K, the time dependence of NMR signal intensity after the saturation pulse (the recovery curve of the nuclear magnetization) can not be fitted to Eq.~(\ref{single}). Instead the recovery curves at peaks A and C can be fitted to the stretched exponential function, 
\begin{equation}\label{stretch}
I(t)=I_{\rm{eq}}\left(1-\exp(-t/T_1)^{\beta}\right), 
\end{equation}
as shown in Fig.~\ref{fig:T1_intermediate2}(b). The deviation of $\beta$ from one provides a measure of the distribution of $1/T_1$. For example, stretched exponential relaxation behavior with $\beta$ = 0.5 has been observed in alloys with dilute magnetic impurities~\cite{McHenry}, where the hyperfine field at a nucleus decays as $1/r^3$ with the distance $r$ from the impurity spin. With decreasing temperature, $(1/T_1)_c$ at peak A and peak C decrease rapidly below about 4~K, similarly to the behavior at higher pressures above 4~GPa.  The stretch exponent $\beta$ also decreases with lowering temperature as shown in Fig. \ref{fig:T1_intermediate2}(c). Although we are not aware of physical models to reproduce stretched exponential behavior with $\beta < 0.5$, the results qualitatively point to the growing inhomogeneity with lowering temperature.  

The recovery curve at peak B also shown in Fig.~\ref{fig:T1_intermediate2}(b), on the other  hand, exhibits two step behavior below 2~K expressed by the sum of two exponential functions with different relaxation rates,
\begin{equation}\label{double_T1}
I(t)=I_{\rm{eq}}-I_f\exp{(-t/T_{1f})}-I_s\exp{(-t/T_{1s})},
\end{equation}
where $1/T_{1f}$ and $1/T_{1s}$ represent the fast and slow relaxation rates.  While $1/T_{1s}$ decreases on cooling in a similar way to the peak A and C, $1/T_{1f}$ keeps large values down to the lowest temperature, indicating persistence of strong spin fluctuations. Thus, even though the NMR spectrum at 3.66~GPa may be understood as a superposition of the spectra from two ordered phases, $1/T_1$ data suggest existence of a distinct paramagnetic region.  

\begin{figure}[t]
\includegraphics[width=8cm]{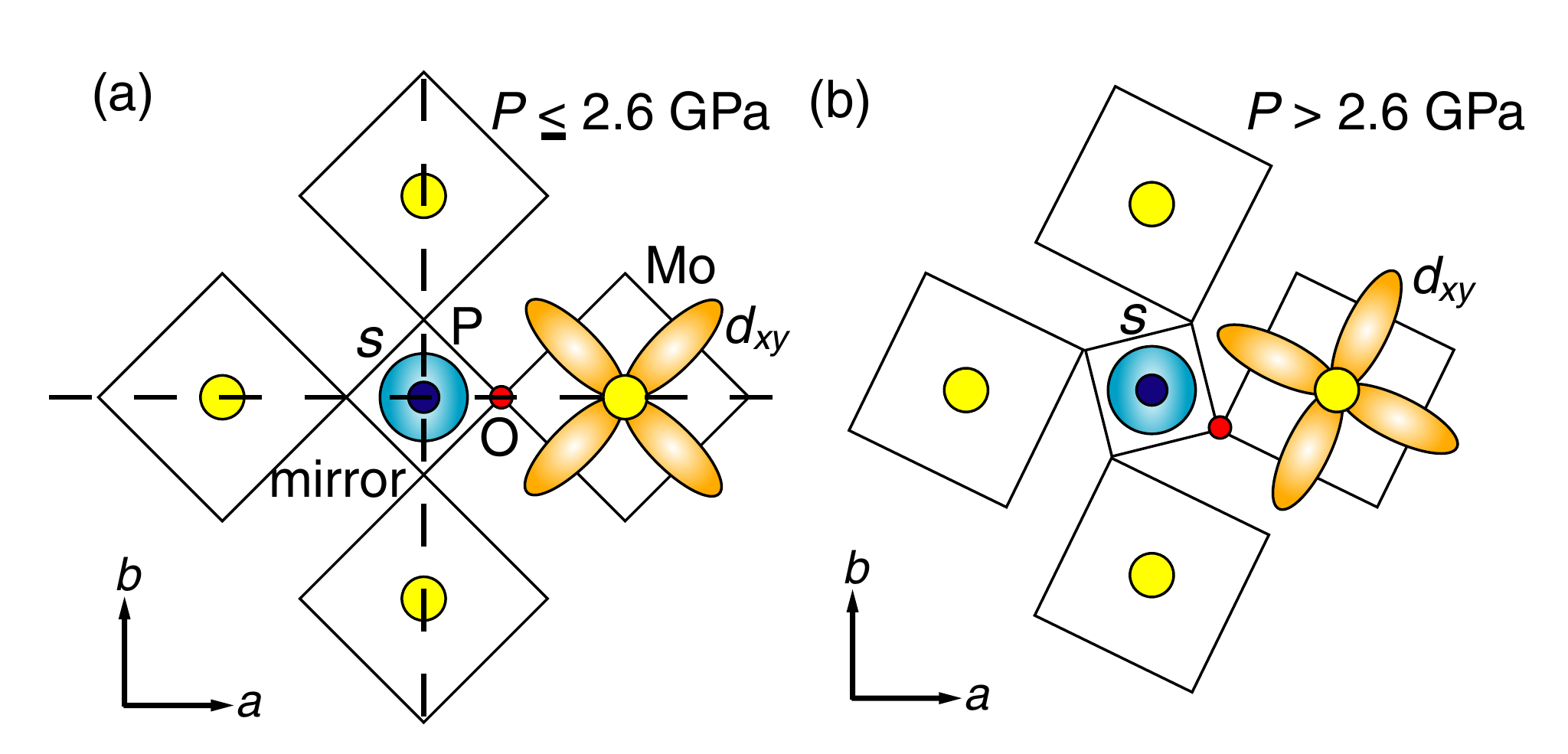}
\caption{\label{fig:orbital} (Color online) (a) Schematic structure of the MoOPO layer in RbMoOPO$_4$Cl at ambient pressure. The Mo-$d_{xy}$ and P-$s$ orbitals have opposite parities with respect to the mirror plane indicated by the dashed lines. Therefore, they do not hybridize and no spin transfer is allowed to the P-$s$ states. (b) Proposed structural distortion at high pressures $P >P_c$, which allows spin transfer to the P-$s$ states.}
\end{figure}

\section{Discussions}\label{discussion}
\subsection{Pressure induced structural transition}
As mentioned in Sec.~\ref{shift}, anomaly in the pressure dependence of $K_c$ (Fig.~\ref{fig:K_P}) strongly suggests a pressure-induced structural transition at $P_c$ = 2.6~GPa. Furthermore, the contrasting behavior of $A_{\rm{iso}}$ and $A_{\rm{ax}}$ above $P_c$ (Fig.~\ref{fig:K_chi_plot}(c)) indicates that the structural distortion above $P_c$ substantially promotes hybridization between Mo-$d$ and P-$s$ orbitals. 

At ambient pressure, the Mo-P bond is on the mirror plane as shown in Fig.~\ref{fig:orbital}(a). Because of the opposite parities of Mo-$d_{xy}$ and P-$s$ orbitals with respect to the mirror, their hybridization is forbidden. Experimentally, a small negative value of $A_{\rm{iso}} = -0.05$ T/$\mu_{\rm{B}}$ is observed in addition to the axial component $A_{\rm{ax}} = -0.03$ T/$\mu_{\rm{B}}$ at ambient pressure (Table~\ref{tab:table1}). This can be understood as the core polarization effect caused by the spin density on P-$p$ orbitals. On the other hand, the large positive value of $A_{\rm{iso}}$ at high pressures above $P_c$ (0.4~T/$\mu_{\rm{B}}$ at 4~GPa) should be ascribed to the spin density on the P-$s$ orbital due to direct hybridization with the Mo-$d_{xy}$ orbital. In order for this to occur, the mirror symmetry must be broken above $P_c$.

This observation reminds us of the structure of MoOPO$_4$, in which MoO$_6$ and PO$_4$ polyhedra are twisted around the $c$-axis in opposite directions, breaking the mirror symmetry as shown in Fig.~\ref{fig:RbMoOPO4Cl_MoOPO4}(c)~\cite{Yang}. This structure allows hybridization between the Mo-$d_{xy}$ and the P-$s$ orbitals [Fig.~\ref{fig:orbital}(b)]. Indeed, NMR experiments on both powder~\cite{Lezama} and single crystal~\cite{Vennemann} samples of MoOPO$_4$ revealed a large positive hyperfine coupling constant at $^{31}$P nuclei. 

The structural distortion in MoOPO$_4$ also has significant influence on magnetic properties. The magnetic properties of MoOPO$_4$ are very different from those of RbMoOPO$_4$Cl at ambient pressure~\cite{Yang}. For MoOPO$_4$, the exchange couplings are estimated as  $J_1$ = 11.4 K and $J_2$ = $-$5.2 K ($J_2/J_1$ = $-$0.46) and NAF-type magnetic order has been confirmed by the neutron scattering experiment~\cite{Yang}. As we discuss in Sec.~\ref{Magnetic structure}, the NAF order is also indicated by the NMR spectra of RbMoOPO$_4$Cl above 3.9~GPa. Therefore, it is likely that the structural transition in RbMoOPO$_4$Cl is accompanied by the twist of MoO$_6$ and PO$_4$ polyhedra similar to MoOPO$_4$.  We should emphasize that while the structural transition takes place sharply at $P_c$ = 2.6~GPa, the change of antiferromagnetic structure occurs over a finite range of higher pressures between 3.4 and 3.8~GPa.  

\subsection{Magnetic structure}\label{Magnetic structure}
\begin{figure}[t]
\includegraphics[width=6cm]{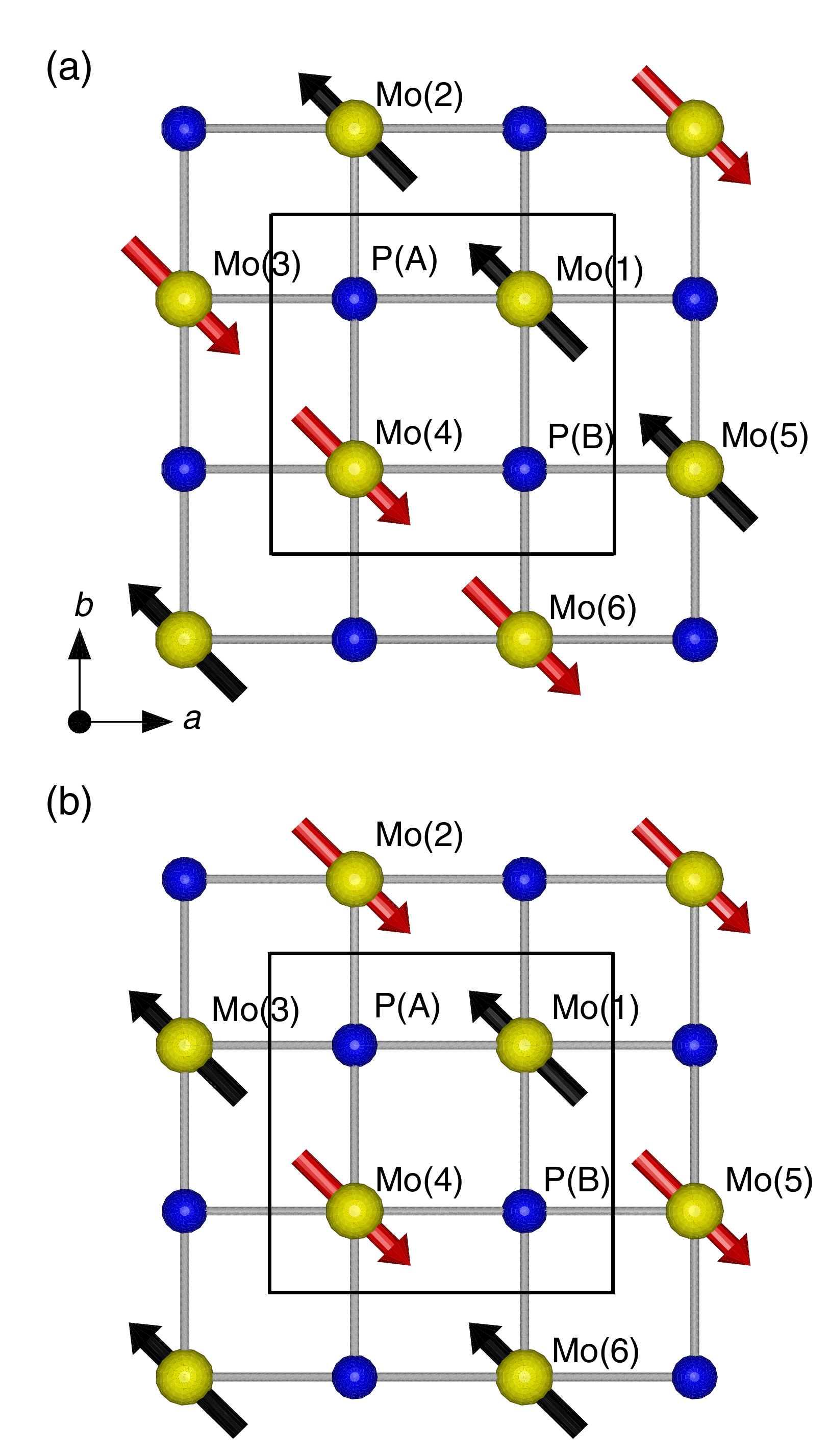}
\caption{\label{fig:RbMoOPO4Cl_magnetic_structure} (Color online)  Schematic structures of (a) CAF and (b) NAF antiferromagnetic order in the MoOPO layer of RbMoOPO$_4$Cl. The solid lines represent the unit cell.}
\end{figure}

Here we discuss the antiferromagnetic (AF) spin structures based on the NMR spectra. We  consider the hyperfine fields at $^{31}$P nuclei generated by two possible ordered structures, CAF and NAF, shown in Fig.~\ref{fig:RbMoOPO4Cl_magnetic_structure}.  In both cases, P sites are divided into two sublattices, P(A) and P(B). The hyperfine fields ${\bf{B}}^{\rm{A}}_{\rm{hf}}$ and ${\bf{B}}^{\rm{B}}_{\rm{hf}}$ at P(A) and P(B) sites can be expressed as 
\begin{equation}
{\bf{B}}_{\rm{hf}}^{\rm{A}}=\sum_{i=1,2,3,4}{\bf{A}}^{\rm{A}}({\bf{r}}_i){\bm{\mu}}_i,
\end{equation}
\begin{equation}
{\bf{B}}_{\rm{hf}}^{\rm{B}}=\sum_{i=1,4,5,6}{\bf{A}}^{\rm{B}}({\bf{r}}_i){\bm{\mu}}_i,
\end{equation}
where ${\bf{A}}^{\rm{A}}({\bf{r}}_i)$ [${\bf{A}}^{\rm{B}}({\bf{r}}_i)$] is the hyperfine coupling tensor between the nucleus at P(A) [P(B)] and the Mo moment ${\bm{\mu}}_i$ at ${\bf{r}}_i $. The ${\bf{A}}^{\rm{A}}({\bf{r}}_i)$ ($i$=1-4) are already given by Eqs. (\ref{A1})-(\ref{A4}). Since P(B) site is related to P(A) by inversion at  $(a/2, a/2, 0)$, ${\bf{A}}^{\rm{B}}({\bf{r}}_i)$ can be obtained from ${\bf{A}}^{\rm{A}}({\bf{r}}_i)$ as ${\bf{A}}^{\rm{B}}({\bf{r}}_1)={\bf{A}}^{\rm{A}}({\bf{r}}_4)$, ${\bf{A}}^{\rm{B}}({\bf{r}}_4)={\bf{A}}^{\rm{A}}({\bf{r}}_1)$, ${\bf{A}}^{\rm{B}}({\bf{r}}_6)={\bf{A}}^{\rm{A}}({\bf{r}}_2)$, ${\bf{A}}^{\rm{B}}({\bf{r}}_5)={\bf{A}}^{\rm{A}}({\bf{r}}_3)$. 

The CAF structure is described as
\begin{eqnarray}
\begin{split}
\braket{{{\bm{\mu}}}_{1}}=\braket{{{\bm{\mu}}}_{2}}=\braket{{{\bm{\mu}}}_{5}}&=-\braket{{{\bm{\mu}}}_{3}}=-\braket{{{\bm{\mu}}}_{4}}=-\braket{{{\bm{\mu}}}_{6}}\\&=\left(\begin{array}{c}
\sigma_a \\
\sigma_b\\
\sigma_c
\end{array}\right).
\label{CAF}
\end{split}
\end{eqnarray}
Then the hyperfine fields $\braket{{\bf{B}}_{\rm{hf}}^{\rm{A}}}$ and $\braket{{\bf{B}}_{\rm{hf}}^{\rm{B}}}$ are expressed as 
\begin{eqnarray}
\begin{split}
\braket{{\bf{B}}_{\rm{hf}}^{\rm{A}}}
&=-\braket{{\bf{B}}_{\rm{hf}}^{\rm{B}}}
\\&=2 \left(\begin{array}{c}
\sigma_c \left(A_{ac}+A_{bc}\right)\\
\sigma_c\left(A_{bc}-A_{ac}\right)\\
\sigma_a\left(A_{cb}+A_{ca}\right)+\sigma_b\left(A_{cb}-A_{ca}\right)
\end{array}\right).
\label{H_AB_CAF}
\end{split}
\end{eqnarray}
This shows that the $c$-component ($ab$-component) of the AF moment produces the hyperfine field in the $ab$-plane (along the $c$-direction). 

The NAF structure is described by  
\begin{eqnarray}
\begin{split}
\braket{{{\bm{\mu}}}_{1}}=\braket{{{\bm{\mu}}}_{3}}=\braket{{{\bm{\mu}}}_{6}}&=-\braket{{{\bm{\mu}}}_{2}}=-\braket{{{\bm{\mu}}}_{4}}=-\braket{{{\bm{\mu}}}_{5}}\\&=\left(\begin{array}{c}
\sigma'_a \\
\sigma'_b\\
\sigma'_c
\end{array}\right),
\label{NAF}
\end{split}
\end{eqnarray}
which produces the hyperfine fields, 
\begin{eqnarray}
\begin{split}
\braket{{\bf{B}}_{\rm{hf}}^{\rm{A}}}
&=-\braket{{\bf{B}}_{\rm{hf}}^{\rm{B}}}
\\&=2\left(\begin{array}{c}
\sigma'_a\left(A_{aa}-A_{bb}\right) +\sigma'_b \left(A_{ab}+A_{ba}\right)\\
\sigma'_a\left(A_{ab}+A_{ba}\right) +\sigma'_b \left(A_{bb}-A_{aa}\right)\\
0
\end{array}\right).
\label{H_AB_NAF}
\end{split}
\end{eqnarray}
In this case, the $c$-component of the hyperfine field is always zero irrespective of the direction of AF moment. For both CAF and NAF structures, the hyperfine fields at P(A) and P(B) sites have the same magnitude and opposite directions.   

The NMR spectra in our experiments are obtained in the magnetic field of 5.0 T either parallel to the $c$-axis or in the $ab$-plane. Then the splitting of resonance lines can be caused only by the component of hyperfine field parallel to the external field. As described in section~\ref{spectra}, the resonance line splits into two peaks for $B_{\rm{ext}}\parallel c$ in the low-pressure region below 3.3~GPa. From Eqs.~(\ref{H_AB_CAF}) and (\ref{H_AB_NAF}), we conclude that such splitting is possible only for the CAF spin structure with a finite $ab$-component of the AF moment ($\sigma_a$ or $\sigma_b$). The absence of splitting for $B_{\rm{ext}}\parallel ab$ then indicates absence of the $c$-component ($\sigma_c$ = 0). This analysis has been already presented in Ref.~\onlinecite{Ishikawa}. 

Since $A_{bc}$ = 0 at ambient pressure as mentioned in section \ref{shift}, the interval between the split peaks for $B_{\rm{ext}}\parallel c$ is obtained from Eq.~(\ref{H_AB_CAF}) as $4|A_{ca}(\sigma_a - \sigma_b)| = 4 \sqrt{2} \sigma |A_{ca} \cos (\theta + \pi/4)|$, where $\sigma_a = \sigma \cos \theta$ and $\sigma_b = \sigma \sin \theta$. The magnitude of the antiferromagnetic moment has been determined from the neutron diffraction experiment as $\sigma$ = 0.53~$\mu_{\rm{B}}$~\cite{Ishikawa}. The interval of the split NMR peaks extrapolates to 0.031~T at $T$ = 0 [Fig.~\ref{fig:spc_pressure_Dependence}(a)]. Then the off-diagonal hyperfine coupling constant $A_{ca}$ can be obtained as
\begin{equation}
A_{ca} = \frac{0.010}{\cos \left( \theta + \pi/4 \right)} \ \ ({\rm{T}}/\mu_{\rm{B}}) .
\label{A_cac}
\end{equation}

In the high pressure region above 3.88~GPa, the NMR spectra show no line splitting for either $B_{\rm{ext}}\parallel c$ or $B_{\rm{ext}}\parallel ab$. These results are compatible only with the NAF order with the AF moment along the $c$-axis ($\sigma'_a = \sigma'_b$ = 0). We should mention that the same AF structure is observed for  MoOPO$_4$ by the neutron diffraction experiments\cite{Yang}. A distinct feature of the NMR spectra in the high-pressure region is the large uniform internal field reaching $B_n$ = 0.4~T at $P$ = 4~GPa (Fig. \ref{fig:spc_1_5K_P_dependence}). This should be ascribed to the weak ferromagnetism shown in Fig.~\ref{fig:MT_tateiwa}(b). In fact, the values of magnetization ($M$ = 0.2~$\mu_{\rm{B}}$/Mo) and the hyperfine coupling constant ($A_{cc}$ = 0.31~T/$\mu_{\rm{B}}$, Table~\ref{tab:table1}) at 4~GPa give the expected uniform internal field as $B_n=4A_{cc}M$ = 0.24~T. This is in reasonable agreement with the experimental value of 0.4~T, considering the different conditions for the  magnetization (at $T$ = 5~K on powder sample) and the NMR (at 1.5 K on a single crystal) experiments.   

In the intermediate pressure range between 3.4 and 3.8~GPa, the NMR spectra show three broad peaks (Fig.~\ref{fig:spc_1_5K_P_dependence}). This spectral shape is naturally understood as a superposition of the spectra from the low-pressure and high-pressure regions and indicates coexistence of the CAF and NAF structures. At 3.66~GPa, for instance, peak A and peak B would be assigned to the CAF phase and peak C to the NAF phase (Fig.~\ref{fig:spc_pressure_Dependence}f). 
Then the variation of NMR spectra with pressure indicates the first-order transition from CAF to NAF spin structure with distribution of the critical pressure over a finite range due to inevitable inhomogeneity or disorder. 

However, the results of $1/T_1$ at 3.66~GPa shown in Fig.~\ref{fig:T1_intermediate2} are not completely consistent with this interpretation. Although $1/T_1$ at peak A and peak C decreases rapidly at low temperatures, consistent with the general behavior in ordered states, the relaxation curve at peak B shows two components of $1/T_1$, one of which exhibits paramagnetic behavior down to 1.5~K as mentioned in section~\ref{spin lattice relaxation rate}.

\subsection{Spin dynamics}\label{T1_discussion}
The spin lattice relaxation rate $1/T_1$ is generally expressed by using the time correlation function of the hyperfine field as~\cite{Moriya1, Moriya2}
\begin{equation}\label{T1_0}
1/T_1=\frac{\gamma^2_{\rm{N}}}{2}\int dt e^{i\omega_{\rm{N}} t}\braket{\{B^+_{\rm{hf}}(t), B^-_{\rm{hf}}(0)\}},
\end{equation}
where $\omega_{\rm{N}}=2\pi\nu$ is the  NMR frequency and $\{A, B\}\equiv (AB+BA)/2$. The transverse components of hyperfine field is defined as
\begin{equation}
B^{\pm}_{\rm{hf}}=B^{\zeta}_{\rm{hf}}\pm iB^{\eta}_{\rm{hf}},
\end{equation}
where $\zeta$ and $\eta$ denote two orthogonal directions perpendicular to ${\bf{B}}_{\rm{ext}}$.

Here we focus on the behavior of $1/T_1$ in the paramagnetic state in the vicinity of $T_{\rm{N}}$, where $1/T_1$ shows a sharp peak when the ground state has the CAF order but no anomaly when the NAF order appears below $T_{\rm{N}}$.  In this temperature range, the critical antiferromagnetic spin fluctuations should have sufficiently long correlation length. Then the contribution to $1/T_1$ from the critical AF fluctuations can be expressed as 
\begin{equation}\label{T1_CAF_CFC}
\left(\frac{1}{T_1}\right)_c=\frac{4\gamma^2}{N}(A^2_{ac}+A^2_{bc})f_{\parallel }^{\rm{CAF}},
\end{equation}
\begin{equation}\label{T1_CAF_CFAB}
\begin{split}
\left(\frac{1}{T_1}\right)_{ab}= \frac{2\gamma^2}{N}\Big[&(A^2_{ac}+A^2_{bc})f_{\parallel}^{\rm{CAF}} \\
&+2(A^2_{ca}+A^2_{cb})f_{\perp}^{\rm{CAF}}\Big],
\end{split}
\end{equation}
when the ordered state has the CAF structure and 
\begin{equation}\label{T1_NAF_CFC}
\left(\frac{1}{T_1}\right)_c=\frac{4\gamma^2}{N}\left[(A_{aa}-A_{bb})^2+(A_{ab}+A_{ba})^2\right]f_{\perp}^{\rm{NAF}},
\end{equation}
\begin{equation}\label{T1_NAF_CFA}
\left(\frac{1}{T_1}\right)_{ab}=\frac{2\gamma^2}{N}\left[(A_{aa}-A_{bb})^2+(A_{ab}+A_{ba})^2\right]f_{\perp}^{\rm{NAF}}, 
\end{equation}
when the NAF order appears below $T_{\rm{N}}$. Here $f_{\parallel (\perp)}^{\rm{CAF} (\rm{NAF})}$ is the low-frequency amplitude of the critical CAF (NAF) spin fluctuations parallel (perpendicular) to the $c$-axis. These expressions are derived in Appendix \ref{T1_TN}.  

The experimental data in Fig.~\ref{fig:T1_aniso} indicate that at ambient pressure and at 2.6~GPa, both $(1/T_1)_{ab}$ and $(1/T_1)_c$ exhibit a divergent peak at $T_{\rm{N}}$. The height of the peak is quite anisotropic at ambient pressure, $(1/T_1)_{ab}/(1/T_1)_c = 2.34$ at $T_{\rm{N}}$ = 8~K but becomes nearly isotropic at 2.6~GPa [see insets of Figs.~\ref{fig:T1_aniso}(a) and (b)]. Such variation of  anisotropy with pressure is compatible only with the critical CAF fluctuations, since Eqs.~(\ref{T1_NAF_CFC}) and (\ref{T1_NAF_CFA}) show that the NAF fluctuations give a universal value of the anisotropy $(1/T_1)_{ab}/(1/T_1)_c$ = 0.5. This observation is consistent with the CAF order in this pressure range concluded from the NMR spectra.

Assuming that the coupling tensor is symmetric ($A_{ac}=A_{ca}$) and considering the mirror symmetry, $A_{bc}=A_{cb}=0$,  at low pressures below 2.6~GPa, the anisotropy of the peak height is expressed for the CAF fluctuations as 
\begin{equation}\label{ratio_T1}
\frac{(1/T_1)_{ab}}{(1/T_1)_c}=\frac{1}{2}+\frac{f_{\perp}({\bf{Q}}_{\rm{CAF}})}{f_{\parallel}({\bf{Q}}_{\rm{CAF}})}.
\end{equation} 
Then the experimental result  at ambient pressure can be explained by the easy plane anisotropy of the critical spin fluctuations $f_{\perp }({\bf{Q}}_{\rm{CAF}})/ f_{\parallel}({\bf{Q}}_{\rm{CAF}})$ = 1.84, which is consistent with the magnetic structure with the AF moment lying in the $ab$-plane as discussed in section~\ref{Magnetic structure}. On the other hand, the nearly isotropic result at 2.6~GPa, $(1/T_1)_{ab}/(1/T_1)_c = 0.84$ at $T_{\rm{N}}$ = 6.8~K, points to the opposite anisotropy $f_{\perp }({\bf{Q}}_{\rm{CAF}})/ f_{\parallel}({\bf{Q}}_{\rm{CAF}}) \sim$ 0.34. The reason for such a reversal of anisotropy with pressure is still an open question.  

At 4.0 GPa and 6.4 GPa,  no critical divergence of $1/T_1$ is observed for both ${\bf{B}}_{\rm{ext}} \parallel c$ and ${\bf{B}}_{\rm{ext}} \parallel ab$ [Figs.~\ref{fig:T1_aniso}(c) and (d)]. These results rule out critical CAF fluctuations since they cause divergence of $(1/T_1)_{ab}$ regardless of the direction of fluctuations [see Eq.~(\ref{T1_CAF_CFAB})]. Furthermore, Eqs.~(\ref{T1_NAF_CFC}) and (\ref{T1_NAF_CFA}) tell us that our results are compatible only with the critical NAF fluctuations confined along the c-axis, i.e. in-plane critical spin fluctuations must be completely absent. These observations are indeed consistent with the NAF order with the AF moment along the $c$-axis as deduced from the NMR spectra in section~\ref{Magnetic structure}. Such a strong Ising anisotropy of spin fluctuations is surprising for a spin-1/2 system, where the single ion anisotropy is absent and anisotropic interactions such as the Dzyaloshinsky-Moriya interaction allowed for the NNN bonds and other anisotropic exchange interaction should be much smaller than the isotropic Heisenberg interactions. 

Nevertheless, even more intriguing anisotropic behavior is observed at ambient pressure in a wide temperature range as shown in Fig.~\ref{fig:T1_aniso}(a). This result is analyzed in detail in Appendix~\ref{T1_ambient}. At ambient pressure, the spin fluctuations have easy-plane anisotropy, opposite to the high-pressure phase. The analysis presented in Appendix~\ref{T1_ambient} shows that the ratio of the $c$-axis to the $ab$-plane spin correlation functions shows nonmonotonic temperature dependence with a broad peak near 100~K (Fig.~\ref{fig:T1_ratio_ambient}). Although we have no idea at moment to explain the mechanism of this anomalous behavior, any anisotropic phenomena should in principle be caused by strong spin-orbit coupling of Mo-4$d$ electrons. The peculiar anisotropic behavior of spin dynamics is a distinct and novel feature of RbMoOPO$_4$Cl, which deserves further investigation.

\subsection{Phase diagram}\label{Phase_diagram}
\begin{figure}[b]
\includegraphics[width=7.5cm]{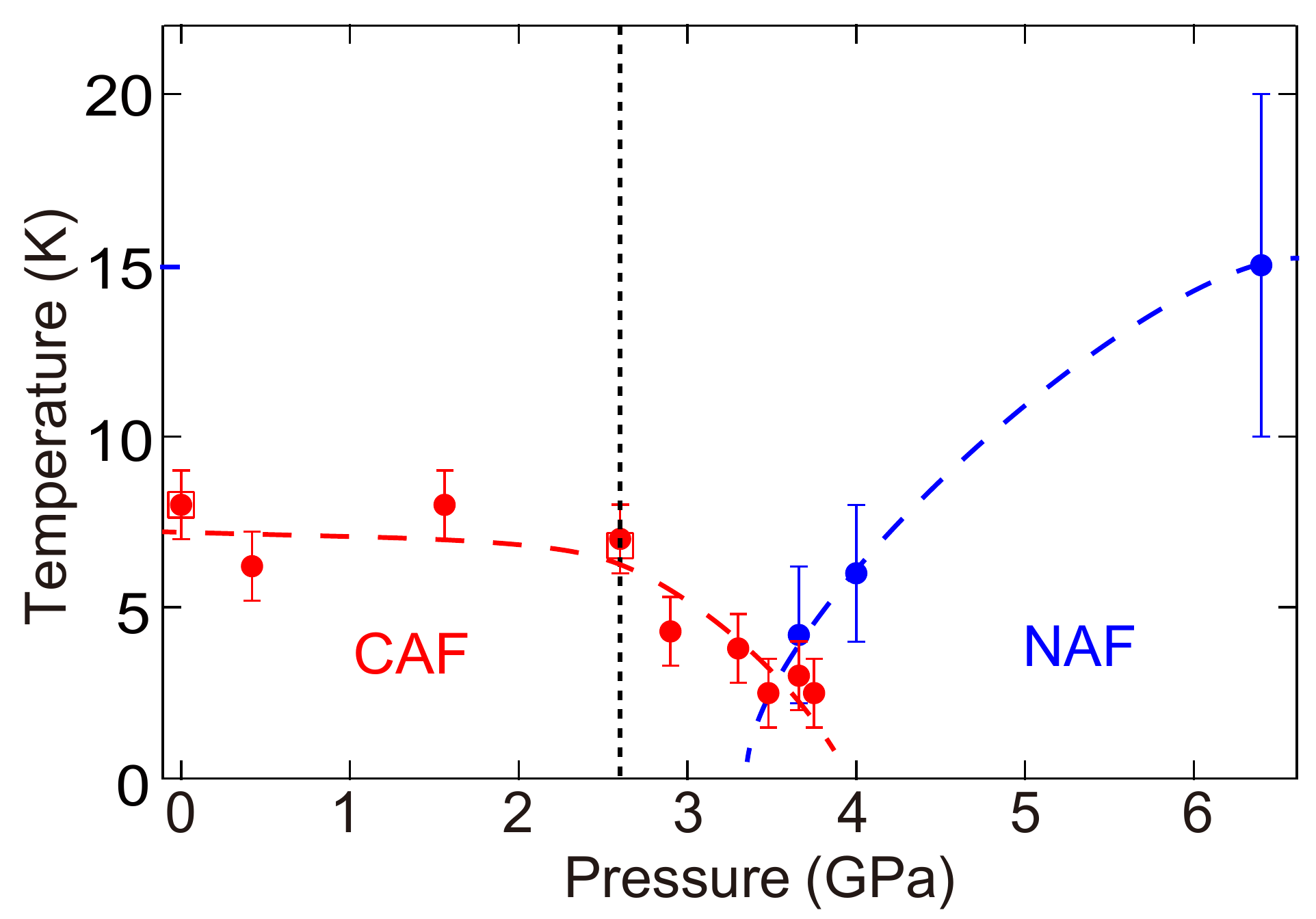}
\caption{\label{fig:PhaseDiagram} (Color online) Pressure vs temperature phase diagram of RbMoOPO$_4$Cl determined by the present $^{31}$P NMR measurements. The dotted vertical line indicates the structural transition at $P_c$ = 2.6~GPa. The red dots (red open squares) represent $T_{\rm{N}}$ of the CAF order determined from the splitting of NMR spectra [the peak of $(1/T_1)_{c}$]. The blue dots indicate $T_{\rm{N}}$ of the CAF order determined from the onset of rapid decrease of $(1/T_1)_c$ marked in Fig.~\ref{fig:T1_intermediate2}. The dashed lines are guides for the eyes.} 
\end{figure}
The pressure versus temperature phase diagram of RbMoOPO$_4$Cl determined from the present NMR measurements is shown in Fig.~\ref{fig:PhaseDiagram}. The vertical dashed line indicates the structural transition at 2.6~GPa identified by the kink in the NMR shift (Fig.~\ref{fig:K_P}). The magnetic transition temperatures $T_{\rm{N}}$ of the CAF phase are determined from the onset of NMR line splitting for $B_{\rm{ext}}\parallel c$ (Fig.~\ref{fig:spc_pressure_Dependence}) or the peak of $(1/T_1)_c$ (Fig.~\ref{fig:T1_aniso}).  Concerning the NAF phase, there is no signature of clear experimental anomaly at $T_{\rm{N}}$. Therefore we determined $T_{\rm{N}}$ from the temperature at which $(1/T_1)_c$ starts to decrease rapidly as marked in Fig.~\ref{fig:T1_intermediate2}, although experimental uncertainty is rather large.    

Below the critical pressure $P_c$ = 2.6~GPa of the structural transition, $J_1$ keeps small values while $J_2$ decreases gradually with pressure as shown in Fig.~\ref{fig:MT_piston}(b). The value of $J_2$ at $P_c$ is about a half of that at ambient pressure. On the other hand, $T_{\rm{N}}$ of the CAF order stays nearly constant 7 $-$ 8~K. This can be understood if the interlayer coupling increases with pressure, compensating for the reduction of $J_2$. In contrast to the modest variation of exchange interactions below $P_c$, much more drastic changes of $J_1$, $J_2$, and their ratio are inferred above $P_c$, given the rapid decrease of $T_{\rm{N}}$ and the change of spin structure from CAF to NAF only slightly above $P_c$, although we do not have sufficient data to determine $J_1$ and $J_2$ quantitatively above $P_c$. We propose that the change of crystal symmetry involving twist of MoO$_6$ and PO$_4$ polyhedra would be the key driving force for this change of behavior. The CAF phase begins to be partially replaced by the NAF phase above 3.3~GPa and completely disappears near 3.9~GPa, above which only the NAF phase survives. The NAF phase has substantially higher $T_{\rm{N}}$ than the CAF phase, 

In the intermediate pressure range between 3.3 and 3.9~GPa, the NMR spectra show coexistence of the CAF and NAF spin structures. This indicates a first-order transition between two ordered phases with distribution of the critical pressure owing to certain inhomogeneity or disorder.  Indeed the direct transition between these two ordered phases has to be first order because of the distinct broken symmetries. However, $1/T_1$ data at 3.66~GPa reveal existence of a paramagnetic component with persistent spin fluctuations down to 1.5~K (Fig.~\ref{fig:T1_intermediate2}). It has been theoretically proposed that spin liquid states with zero or a small energy gap may exist between the two ordered phases~\cite{Sushkov, Zhang, Capriotti1, Singh, Sirker, Capriotti2, Jiang, Mezzacapo, Gong}.  Thus the $1/T_1$ data raises possibility of a spin liquid phase between the CAF and NAF phases which, however, may be fragile against disorder and difficult to identify with spectroscopic evidence. Whether the paramagnetic behavior revealed by the $1/T_1$ measurements is an intrinsic signature of a quantum disordered phase or caused by extrinsic disorder is still an open question. Further investigations including experiments at lower temperatures are required.

\section{Conclusion}
We performed magnetization and $^{31}$P NMR measurements on RbMoOPO$_4$Cl under high pressures. The large enhancement of magnetization with pressure indicates that high pressure is a very efficient tool to tune exchange interactions in this material. This is also evidenced by the strong pressure dependence of hyperfine coupling constants, which are determined by the orbital hybridization between Mo and P sites. A pressure-induced structural phase transition at $P_c$ = 2.6~GPa is revealed by the anomaly in the pressure dependence of the NMR shift. The substantial increase of the isotropic hyperfine coupling constant above $P_c$ suggests breaking of mirror symmetry, likely resulting from the twist of MoO$_5$Cl and PO$_4$ polyhedra similar to what was observed in MoOPO$_4$.

The structural transition accelerates the variation of exchange interactions with pressure and drives the magnetic phase transition at higher pressure region. The pressure dependence of the NMR spectra at low temperatures reveals the change of antiferromagnetic spin structure from CAF to NAF, which takes place over a finite range of pressure 3.3$-$3.9~GPa. The nuclear spin lattice relaxation rate $1/T_1$ exhibits a divergent peak at $T_{\rm{N}}$ at low pressures but such a peak is absent at high pressures. This result can be also explained by the change of critical spin fluctuations from CAF type to NAF type. 

In the intermediate pressure region between 3.3 and 3.9~GPa, the NMR spectra indicate coexistence of the CAF and NAF phases, pointing to a first-order transition influenced by disorder. However, a certain component of $1/T_1$ at $P$ = 3.66~GPa shows paramagnetic behavior with persistent spin fluctuations down to the lowest temperature. Whether this behavior is an intrinsic signature of a quantum disordered phase or not is still an open question..  

Finally, we remark on the highly anisotropic behavior unexpected for a spin 1/2 system.  At ambient or low pressures, both the CAF ordered moment and the spin fluctuations show easy plane anisotropy. Furthermore, the anisotropy of spin fluctuations at ambient pressure exhibits very unusual nonmonotonic temperature dependence. The anisotropy gets reversed at higher pressures. Both the NAF ordered moment and the critical spin fluctuations near $T_{\rm{N}}$ show strong Ising anisotropy. The puzzling anisotropic behavior appears to be a distinct feature of RbMoOPO$_4$Cl, likely to be caused by strong spin orbit interaction of Mo-$4d$ electrons.  

\begin{acknowledgments}
The authors would like to thank M. Yoshida and N. Shannon for fruitful discussions.  This study was supported by the JSPS KAKENHI Grant Nos. JP17H02918, JP25287083, JP18H04310 (J-Physics), JP17K05531, JP20K03829, and JP20K03830.

\end{acknowledgments}

\appendix
\section {Analysis of the SQUID data under high pressures}\label{MT4GPa}
\begin{figure}[t]
\includegraphics[width=8cm]{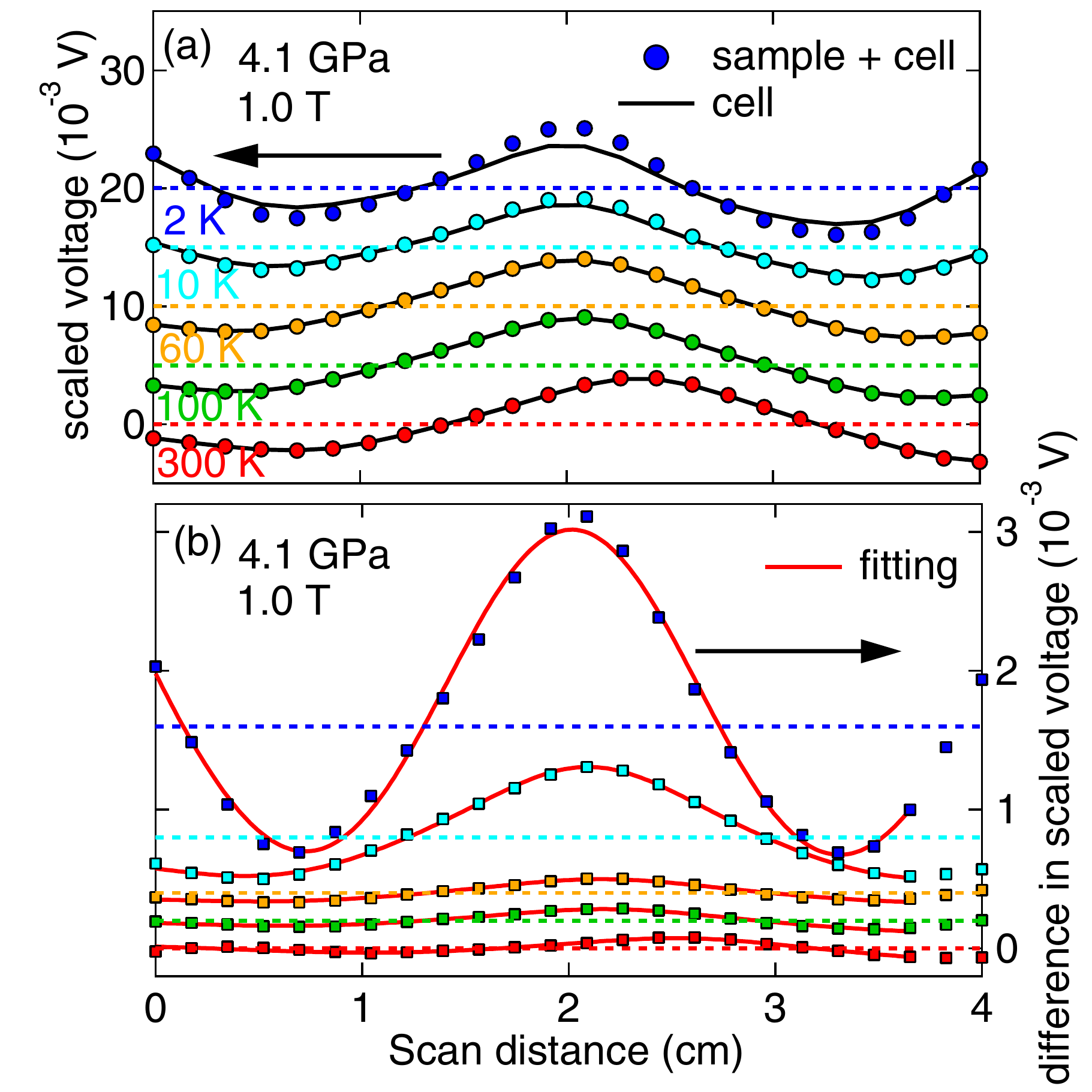}
\caption{\label{fig: tateiwa_4GPa_curve} (Color online) (a) The SQUID response obtained from the standard 4~cm long DC scan.  The circles represent the data obtained from the pressure cell with the sample at 4.1~GPa and $B_{\rm{ext}}$ = 1.0 T in the temperature range 2~K $\leq T \leq$ 300~K. The solid lines represent the data from the cell without the sample and the dashed lines are the base lines for each temperature.  (b) The difference of the SQUID responses with and without the sample.  The solid lines show the fitting to Eq.~(\ref{MT_squid}).}
\end{figure}

We describe the procedure to measure magnetization under high pressures.  Figure \ref{fig: tateiwa_4GPa_curve}(a) shows the SQUID response obtained from the standard 4~cm long DC scan at  $B_{\rm{ext}}$ = 1.0 T under the pressure of 4.1~GPa at various temperatures. The circles represent the data from the pressure cell with the sample, while the solid lines shows the data from the cell without the sample. Their differences are plotted in  Fig.~\ref{fig: tateiwa_4GPa_curve}(b) and fitted to the function,
\begin{equation}
\begin{split}{\label{MT_squid}}
f(Z)= &a_1+a_2Z+a_3\Big[\frac{2}{[R^2+(Z+a_4)^2]^{3/2}} \\
&-\frac{1}{[R^2+(\Gamma+(Z+a_4))^2]^{3/2}} \\
&-\frac{1}{[R^2+(-\Gamma+(Z+a_4))^2]^{3/2}}\Big],
\end{split}
\end{equation}
where $Z$ is the scan distance and $a_1$, $a_2$, $a_3$, and $a_4$ are the fitting parameters. The constants $\Gamma$ and $R$ are related to the geometry of the detection coil of SQUID and known in advance. The magnetization is obtained from the value of  $a_3$ by multiplying a conversion factor~\cite{Tateiwa, QD}. 

At low temperatures below 10~K, the two sets of SQUID data with and without the sample can be clearly distinguished in the raw data in Fig.~\ref{fig: tateiwa_4GPa_curve}(a) and their difference displayed in Fig.~\ref{fig: tateiwa_4GPa_curve}(b) shows a peak at the right position. Thus we are confident that the contribution from the sample is correctly obtained by this procedure. On the other hand, at high temperatures, the two sets of SQUID data are almost indistinguishable in Fig.~\ref{fig: tateiwa_4GPa_curve}(a) and the peak position of the difference curve in Fig.~\ref{fig: tateiwa_4GPa_curve}(b) is significantly shifted from the expected sample position. Therefore most likely the difference curves are heavily influenced by the inevitable irreproducibility of the SQUID response from the pressure cell in addition to the contribution from the sample. We believe that this is the primary reason for the large deviation above 100~K between the susceptibility data at high pressures and the data at ambient pressure measured without the pressure cell displayed in Fig.~\ref{fig:MT_tateiwa}(a).    

\section {Axial symmetry of the shift at 6.4 GPa}\label{6.4GPa}
\begin{figure}[t]
\includegraphics[width=8cm]{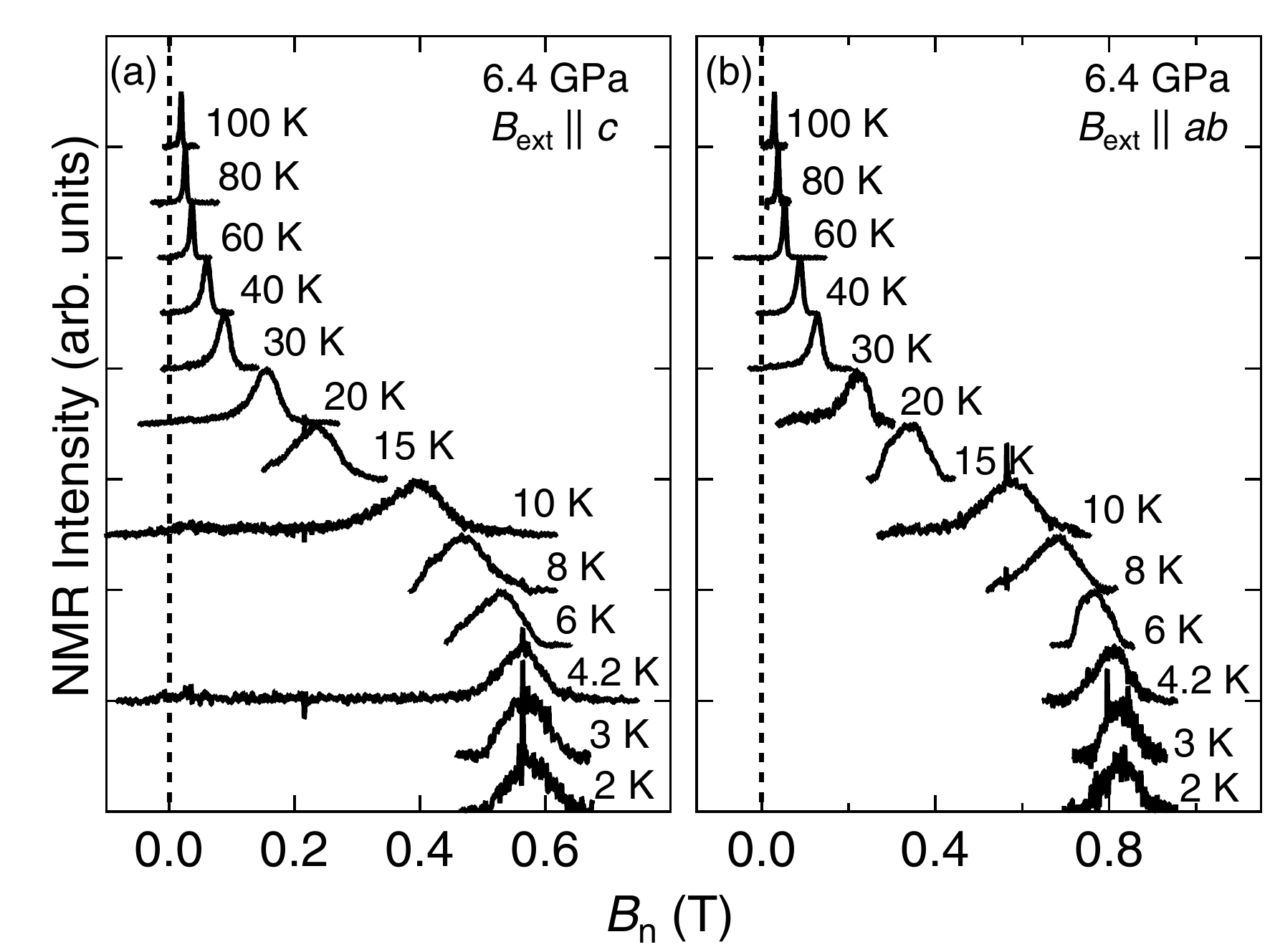}
\caption{\label{fig:spc_6_4GPa} Temperature dependence of the $^{31}$P NMR spectra at 6.4~GPa with the magnetic field of 5.0 T (a) along the $c$-axis and (b) in the $ab$-plane.}
\end{figure}

\begin{figure}[t]
\includegraphics[width=8cm]{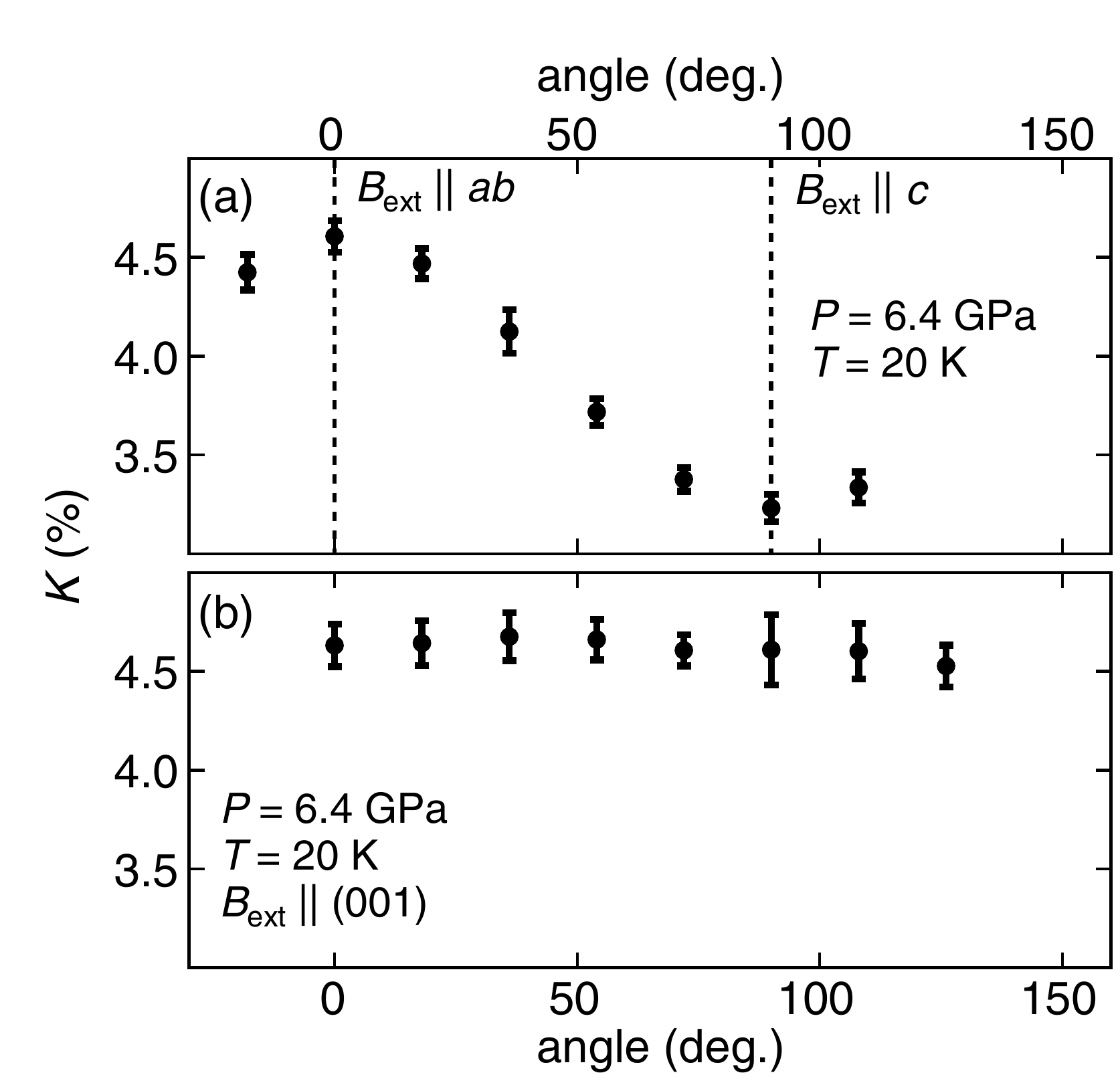}
\caption{\label{fig:K_6_4GPa} Angle dependence of NMR shift $K$ measured at 20~K and 6.4~GPa with the magnetic field (a) tilted from the $c$-axis to the $ab$-plane and (b) applied in the $ab$-plane.}
\end{figure}

Figures \ref{fig:spc_6_4GPa}(a) and \ref{fig:spc_6_4GPa}(b) show temperature dependences of the $^{31}$P NMR spectra at 6.4 GPa for ${\bf{B}}_{\rm{ext}}\parallel c$ and ${\bf{B}}_{\rm{ext}} \parallel ab$.  As is the case at 4.0 GPa shown in Figs. \ref{fig:spc_pressure_Dependence}(h) and \ref{fig:spc_pressure_Dependence}(j), the spectra shifts to larger values of $B_{\rm{n}}$ with decreasing temperature without splitting at low temperatures. Figures~\ref{fig:K_6_4GPa}(a) and \ref{fig:K_6_4GPa}(b) show the angle dependence of $K$ at $T$ = 20~K. When the field is tilted from the $c$-axis to the $ab$-plane, the angle dependence is described as $K(\theta) = K_c \cos^2\theta + K_{ab} \sin^2\theta$, which is the same relation observed at ambient pressure. On the other hand, when the field is rotated in the $ab$ plane, $K$ stays constant within the experimental error, i.e., K is isotropic in the $ab$ plane. This indicates that the $S_4$ symmetry at P site is preserved under high pressure.  

{\section {Spin dynamics in the vicinity of $T_{\rm{N}}$}\label{T1_TN}
Here we derive the expressions Eqs.~(\ref{T1_CAF_CFC})$-$(\ref{T1_NAF_CFA}) for the contribution to $1/T_1$ from critical AF spin fluctuations.
By using the Fourier transform of the hyperfine coupling tensor ${\bf{A}}({\bf{q}})$ and magnetic moment ${\bm{\mu}}({\bf{q}})$ for RbMoOPO$_4$Cl defined in the two dimensional ${\bf{q}}$ space as 
\begin{equation}\label{Aq}
{\bf{A}}({\bf{q}})=\sum_{i=1}^4 {\bf{A}}({\bf{r}}_i) e^{i{\bf{q}}\cdot{\bf{r}}_i},
\end{equation}
\begin{equation}\label{mu_q}
{\bm{\mu}}({\bf{q}})=\frac{1}{\sqrt{N}}\sum_i {\bm{\mu}}({\bf{r}}_i) e^{-i{\bf{q}}\cdot{\bf{r}}_i}.
\end{equation}
The hyperfine field ${\bf{B}}_{\rm{hf}}$ can be written as 
\begin{equation}\label{Bq}
{\bf{B}}_{\rm{hf}}=\frac{1}{\sqrt{N}}\sum_{\bf{q}} {\bf{A}}({\bf{q}}){\bm{\mu}}({\bf{q}}).
\end{equation}

Let us consider the relaxation rate at the P(A) site in Fig.~\ref{fig:RbMoOPO4Cl_MoOPO4}.  By substituting Eqs.~(\ref{A1})-(\ref{A4}) into Eq.~(\ref{Aq}), $1/T_1$ can be expressed as,
\begin{equation}\label{T1_q}
\frac{1}{T_1}=\frac{\gamma^2}{2N}\sum_{\bf{q}} \sum_\alpha \left[|A_{\zeta\alpha}({\bf{q}})|^2+|A_{\eta\alpha}({\bf{q}})|^2\right]\mu_{\alpha\alpha}({\bf{q}}),
\end{equation}
where $A_{\zeta \alpha}({\bf{q}})$ are the $\zeta \alpha$-component of the hyperfine form factor tensor ${\bf{A}}({\bf{q}})$ given by 
\begin{widetext}
\begin{eqnarray}
{{\bf{A}}}({\bf{q}})&=2\left(\begin{array}{ccc}
A_{aa}\cos{\frac{q_a}{2}}+A_{bb}\cos{\frac{q_b}{2}} & A_{ab}\cos{\frac{q_a}{2}}-A_{ba}\cos{\frac{q_b}{2}} & i\left(A_{ac}\sin{\frac{q_a}{2}}+A_{bc}\sin{\frac{q_b}{2}}\right)\\
A_{ba}\cos{\frac{q_a}{2}}-A_{ab}\cos{\frac{q_b}{2}} & A_{bb}\cos{\frac{q_a}{2}}+A_{aa}\cos{\frac{q_b}{2}} & i\left(A_{bc}\sin{\frac{q_a}{2}}-A_{ac}\sin{\frac{q_b}{2}}\right)\\
i\left(A_{ca}\sin{\frac{q_a}{2}}+A_{cb}\sin{\frac{q_b}{2}}\right) & i\left(A_{cb}\sin{\frac{q_a}{2}}-A_{ca}\sin{\frac{q_b}{2}}\right) & A_{cc}\left(\cos{\frac{q_a}{2}}+\cos{\frac{q_b}{2}}\right)
\end{array}\right),{\label{Aq_matrix}}
\end{eqnarray}
\end{widetext}
and $\mu_{\alpha\alpha}({\bf{q}})$ ($\alpha=a, b, c$) are dynamical correlation functions of the magnetic moments,
\begin{equation}\label{correlation function}
\mu_{\alpha \alpha}({\bf{q}})=\int dt e^{i\omega_{\rm{N}}t} \braket{\{\mu_{\alpha}({\bf{q}}, t), \mu_{\alpha}(-{\bf{q}},0)\}}.
\end{equation}}

In the vicinity of $T_{\rm{N}}$, where the antiferromagnetic correlation length is sufficiently large, the dominant magnetic fluctuations are confined in a narrow region of ${\bf{q}}$-space in the neighborhood of the ordering wave vector ${\bf{Q}}$.  Then, $1/T_1$ can be expressed as
\begin{equation}\label{T1_q2}
\frac{1}{T_1}=\frac{\gamma^2}{2N}\sum_\alpha \left[|A_{\zeta\alpha}({\bf{Q}})|^2+|A_{\eta\alpha}({\bf{Q}})|^2\right]f_{\alpha\alpha}({\bf{Q}}),
\end{equation}
where $f_{\alpha\alpha}({\bf{Q}})$ is the sum of dynamical correlation functions of the magnetic moments $\mu_{\alpha\alpha}({\bf{q}})$ in ${\bf{q}}$-space around the peak at ${\bf{Q}}$,
\begin{equation}\label{f_q}
f_{\alpha\alpha }({\bf{Q}})=\sum_{\bf{q}} \mu_{\alpha\alpha}({\bf{Q}}+{\bf{q}}).
\end{equation}
In the following, we assume axial anisotropy for the critical spin fluctuations, $f_{aa}({\bf{Q}}) = f_{bb}({\bf{Q}}) \equiv f_{\perp}({\bf{Q}})$ and $f_{cc}({\bf{Q}}) \equiv f_{\parallel}({\bf{Q}})$.

For the CAF structure shown in Fig.~\ref{fig:RbMoOPO4Cl_magnetic_structure}(a) with the spin stripe along [1$\overline{1}$0], the ordering wave vector is given by ${\bf{Q}}_{\rm{CAF,1}}=(\pi, \pi)$. There is  another equivalent CAF domain with the stripe along [110], which is described by ${\bf{Q}}_{\rm{CAF,2}}=(\pi, -\pi)$. The hyperfine coupling tensors for the first domain is given as  
\begin{eqnarray}
{{\bf{A}}}({\bf{Q}}_{\rm{CAF,1}})&=2i\left(\begin{array}{ccc}
0 & 0 & A_{ac}+A_{bc}\\
0 & 0 & A_{bc}-A_{ac} \\
A_{ca}+A_{cb} &A_{cb}-A_{ca} & 0
\end{array}\right), {\label{AQC_matrix}}
\end{eqnarray}
By substituting Eq.~(\ref{AQC_matrix}) into Eq.~(\ref{T1_q2}), we obtain Eq.~(\ref{T1_CAF_CFC}) for $(1/T_1)_c$. We also obtain 
\begin{equation}\label{T1_CAF_CFA1}
\begin{split}
\left(\frac{1}{T_1}\right)_a= \frac{2\gamma^2}{N}\Big[&(A_{ac}-A_{bc})^2f_{\parallel }({\bf{Q}}_{\rm{CAF, 1}}) \\
&+2(A^2_{ca}+A^2_{cb})f_{\perp}({\bf{Q}}_{\rm{CAF, 1}})\Big], \\
\left(\frac{1}{T_1}\right)_b= \frac{2\gamma^2}{N}\Big[&(A_{ac}+A_{bc})^2f_{\parallel }({\bf{Q}}_{\rm{CAF, 1}}) \\
&+2(A^2_{ca}+A^2_{cb})f_{\perp}({\bf{Q}}_{\rm{CAF, 1}})\Big],
\end{split}
\end{equation}
This anisotropy of $1/T_1$ in the $ab$-plane is caused by selecting the first CAF domain. In the paramagnetic state, the critical fluctuations of the second CAF domain should equally contribute to $1/T_1$, for which the expressions for $(1/T_1)_a$ and $(1/T_1)_b$ are exchanged. Therefore, experimentally observed $(1/T_1)_{ab}$ should be the average of the two, leading to the expression in Eq. (\ref{T1_CAF_CFAB}).
Since ${\bf{Q}}_{\rm{CAF,1}}$ and ${\bf{Q}}_{\rm{CAF,2}}$ are equivalent, we set $f_{\alpha\alpha}({\bf{Q}}_{\rm{CAF,1}}) = f_{\alpha\alpha}({\bf{Q}}_{\rm{CAF,2}}) = f_{\alpha\alpha}^{\rm{CAF}}$.

For the NAF structure shown in Fig.~\ref{fig:RbMoOPO4Cl_magnetic_structure}(b), the ordering wave vector is given by ${\bf{Q}}_{\rm{NAF}}=(2\pi, 0)$ and  
\begin{eqnarray}
{{\bf{A}}}({\bf{Q}}_{\rm{NAF}})&=-2\left(\begin{array}{ccc}
A_{aa}-A_{bb} & A_{ab}+A_{ba} & 0 \\
A_{ab}+A_{ba} & A_{bb}-A_{aa} & 0 \\
0 & 0 & 0
\end{array}\right). {\label{AQN_matrix}}
\end{eqnarray}
We then obtain Eqs. (\ref{T1_NAF_CFC}) and (\ref{T1_NAF_CFA}) by substituting Eq.~(\ref{AQN_matrix}) into Eq.~(\ref{T1_q2}) and simplifying the notation as $f_{\perp}^{\rm{NAF}} \equiv f_{\perp}({\bf{Q}}_{\rm{NAF}})$.

\section {Spin dynamics at ambient pressure}\label{T1_ambient}
Here we discuss possible mechanism causing the very unusual temperature dependences of $(1/T_1)_{c}$ and $(1/T_1)_{ab}$ at ambient pressure shown in Fig.~\ref{fig:T1_aniso}(a). First we rewrite the formula for $1/T_1$ in terms of pair correlation functions in real space to analyze spin fluctuations in a wide temperature range far above $T_{\rm{N}}$.  From Eq.~(\ref{hfcoupling}), the correlation function of the hyperfine field is expressed as 
\begin{equation}
\begin{split}
&\int dt e^{i\omega_{\rm{N}} t}\braket{\{B^{\zeta}_{\rm{hf}}(t), B^{\zeta}_{\rm{hf}}(0)\}} \\
&=\sum_{i=1-4}\sum_{j=1-4}\sum_{\alpha=a,b,c}A_{\zeta\alpha}({\bf{r}}_i)A_{\zeta\alpha}({\bf{r}}_j)\mu^{ij}_{\alpha\alpha},
\end{split}
\end{equation}
where
\begin{equation}
\mu^{ij}_{\alpha \alpha}=\int dt e^{i\omega_{\rm{N}}t}\braket{\left\{\mu_{\alpha}({\bf{r}}_i,t),\mu_{\alpha}({\bf{r}}_j, 0)\right\}}.
\end{equation}
is the pair correlation function between the moments at ${\bf{r}}_i$ and ${\bf{r}}_j$. 
In addition to the on-site auto correlation function $\mu_{\alpha \alpha}^{\rm{auto}}=\mu^{ii}_{\alpha \alpha}$, we consider three types of pair correlation among four Mo sites shown in Fig.~\ref{fig:RbMoOPO4Cl_MoOPO4}: two types of the nearest-neighbor correlation, $\mu_{\alpha \alpha}^{\rm{NNA}}$ for the pairs Mo(1)-Mo(2) and Mo(3)-Mo(4) and $\mu_{\alpha \alpha}^{\rm{NNB}}$ for the pairs Mo(1)-Mo(4) and Mo(2)-Mo(3), and the next-nearest-neighbor correlation $\mu_{\alpha \alpha}^{\rm{NNN}}$ for the pairs Mo(1)-Mo(3) and Mo(2)-Mo(4). At ambient pressure, the CAF-type short-range correlation should develop at low temperatures, where $\mu_{\alpha \alpha}^{\rm{NNA}}=-\mu_{\alpha \alpha}^{\rm{NNB}}$ and $\mu_{\alpha \alpha}^{\rm{NNN}} < 0$ [Fig.~\ref{fig:RbMoOPO4Cl_magnetic_structure}(a)].   

In order to simplify the analysis, we assume that all correlation functions have uniaxial anisotropy,
$\mu_{aa}^{\delta}=\mu_{bb}^{\delta}=\mu_{\perp}^{\delta}$, and $\mu_{cc}^{\delta}=\mu_{\parallel }^{\delta}$ ($\delta$= auto, NNA, NNB, or NNN). At ambient pressure, $A_{ab}=A_{ba}=A_{bc}=A_{cb}=0$ because of the mirror symmetry and we assume $A_{ac}=A_{ca}$. Then $(1/T_1)_{ab}$ and $(1/T_1)_c$ can be expressed as 
\begin{equation}
\begin{split}{\label{T1a_pc}}
\left(\frac{1}{T_1}\right)_{ab}&=\gamma^2_{\rm{N}}[(2A^2_{cc}+A^2_{ac})\mu_{\parallel }^{\rm{auto}}+(A^2_{aa}+A^2_{bb}+2A^2_{ac})\mu_{\perp}^{\rm{auto}} \\
&+2A^2_{cc}(\mu_{\parallel }^{\rm{NNA}}+\mu_{\parallel }^{\rm{NNB}})+2A_{aa}A_{bb}(\mu_{\perp}^{\rm{NNA}}+\mu_{\perp}^{\rm{NNB}}) \\
&+(2A^2_{cc}-A^2_{ac})\mu_{\parallel }^{\rm{NNN}}+(A^2_{aa}+A^2_{bb}-2A^2_{ac})\mu_{\perp}^{\rm{NNN}}],
\end{split}
\end{equation}
\begin{equation}
\begin{split}{\label{T1c_pc}}
\left(\frac{1}{T_1}\right)_c&=2\gamma^2_{\rm{N}}[A^2_{ac}\mu_{\parallel }^{\rm{auto}}+(A^2_{aa}+A^2_{bb})\mu_{\perp}^{\rm{auto}} \\
&+2A_{aa}A_{bb}(\mu_{\perp}^{\rm{NNA}}+\mu_{\perp}^{\rm{NNB}}) \\
&-A^2_{ac}\mu_{\parallel }^{\rm{NNN}}+(A^2_{aa}+A^2_{bb})\mu_{\perp}^{\rm{NNN}}],
\end{split}
\end{equation}

We first reexamine the critical behavior of $1/T_1$ near $T_{\rm{N}}$ and demonstrate that the pair correlation formalism leads to the identical results as obtained in section~\ref{T1_discussion}. The critical CAF spin fluctuations with sufficiently large correlation length can be described by the following relations for pair correlation functions,
\begin{equation}{\label{CAF_CF}}
\mu_{\alpha \alpha}^{\rm{auto}}=\mu_{\alpha \alpha}^{\rm{NNA}}=-\mu_{\alpha \alpha}^{\rm{NNB}}=-\mu_{\alpha \alpha}^{\rm{NNN}}=\mu_{\alpha \alpha}^{\rm{cf}}, 
\end{equation}
Basically, $\mu_{\alpha \alpha}^{\rm{cf}}$ and $f_{\alpha\alpha}({\bf{Q}}_{\rm{CAF}})$ in Eq.~(\ref{f_q}) represent the same critical fluctuations. 
By substituting Eq.(\ref{CAF_CF}) into Eqs.(\ref{T1a_pc}) and (\ref{T1c_pc}), $(1/T_1)_{ab}$ and $(1/T_1)_c$ near $T_{\rm{N}}$ can be expressed as
\begin{equation}
\left(\frac{1}{T_1}\right)_{ab}=2\gamma^2_{\rm{N}}\left(2A_{ca}^2\mu_{\perp}^{\rm{cf}}+A^2_{ac}\mu_{\parallel }^{\rm{cf}}\right),
\end{equation}
\begin{equation}
\left(\frac{1}{T_1}\right)_c=4\gamma^2_{\rm{N}}A^2_{ac}\mu_{\parallel }^{\rm{cf}}.
\end{equation}
These are exactly the same results as Eqs.~(\ref{T1_CAF_CFC}) and (\ref{T1_CAF_CFAB}) with $A_{bc}=A_{cb}=0$. 

Let us now examine the temperature dependence and the anisotropy of $1/T_1$ in a wide temperature range below 300~K shown in Fig.~\ref{fig:T1_aniso}(a).  The experimental results are anomalous in two aspects. First, both $(1/T_1)_{c}$ and $(1/T_1)_{ab}$ are strongly temperature dependent even near the highest temperature (300~K) of our measurements, which is ten times larger than $J_2$. Second, the temperature dependences of $(1/T_1)_{c}$ and $(1/T_1)_{ab}$ are remarkably different. This means that not only the auto and pair correlation functions have anomalous temperature dependences but also their anisotropy has peculiar temperature dependence. In order to disentangle these two effects, we make further simplifications as follows. 
\begin{figure}[t]
\includegraphics[width=8cm]{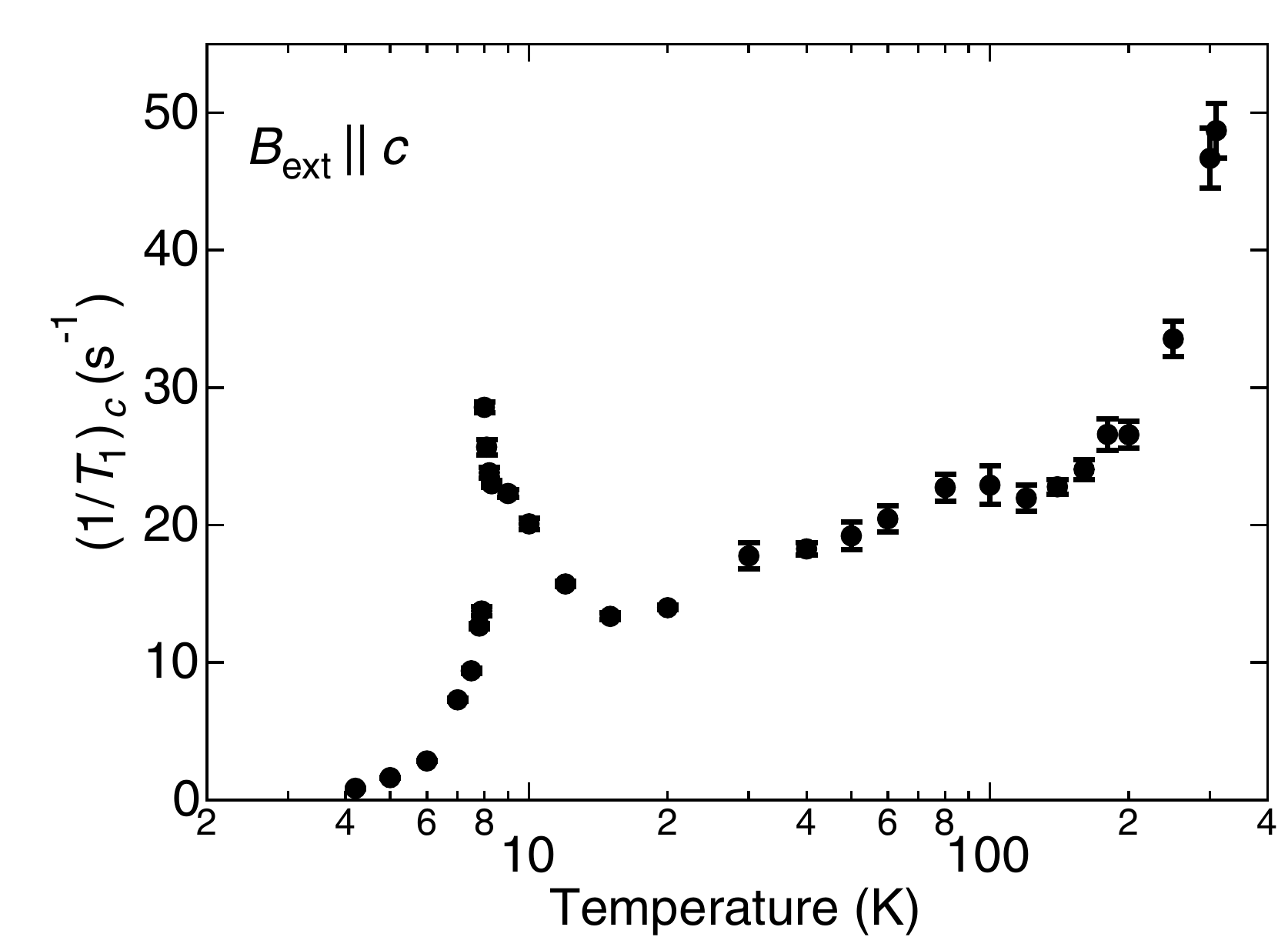}
\caption{\label{fig:T1_c_ambient} Temperature dependence of $(1/T_1)_{c}$ at ambient pressure. The same data as shown in Fig.~\ref{fig:T1_aniso}(a) are plotted here with enlarged vertical scale.}
\end{figure}
\begin{figure}[t]
\includegraphics[width=8cm]{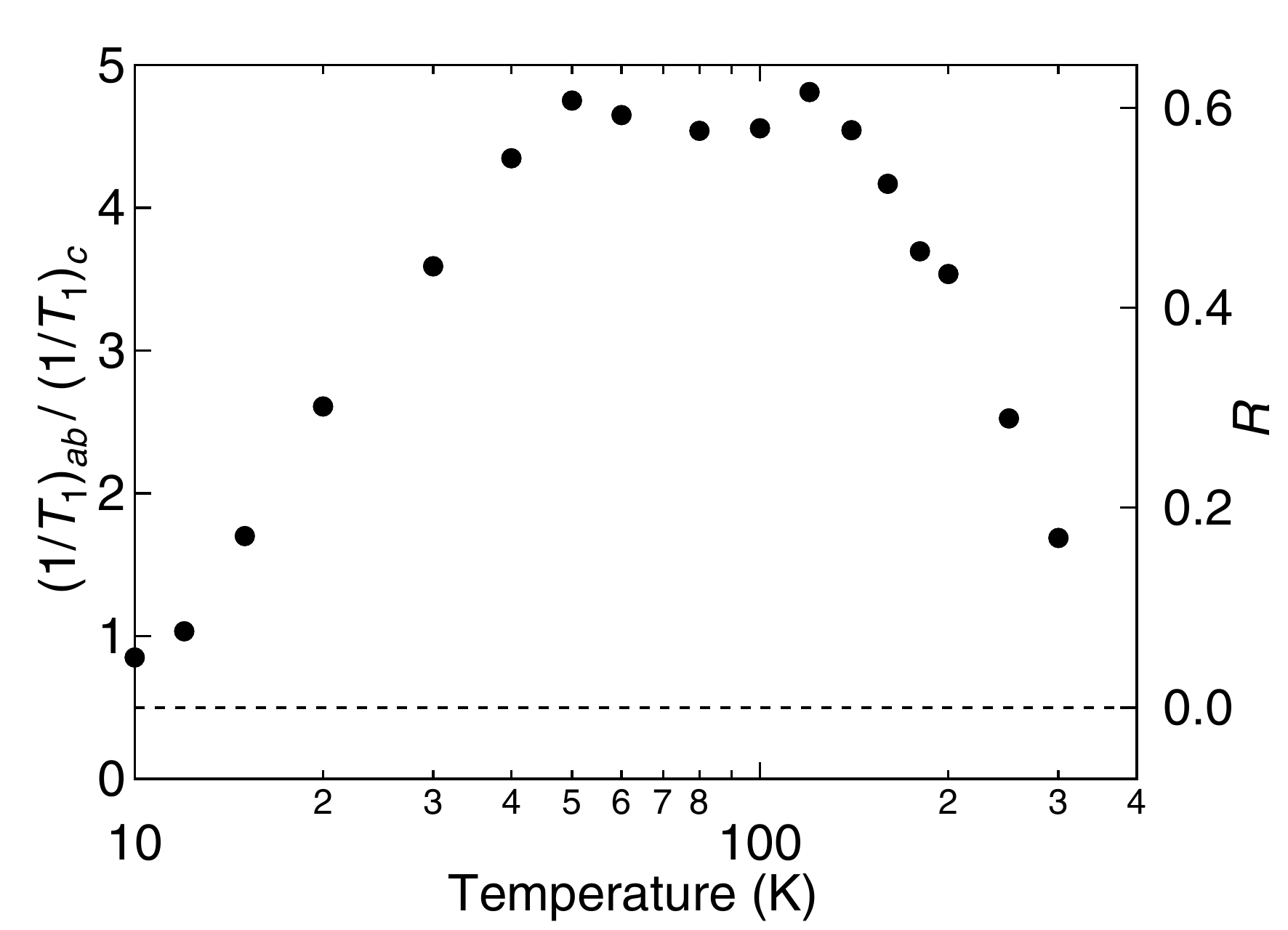}
\caption{\label{fig:T1_ratio_ambient} Temperature dependence of the ratio $(1/T_1)_{ab}/(1/T_1)_{c}$ at ambient pressure indicated by the left axis. The right axis shows the value of $R$ defined by Eq.~(\ref{R}) and obtained from Eq.~(\ref{T1_ratio}) by assuming $\Gamma_2/\gamma^2_{\rm{N}}$ = 0.00146~(T/$\mu_{\rm{B}})^2$.}
\end{figure}

First, we assume that the sum of nearest-neighbor correlations is negligible at ambient pressure, $\mu_{\alpha\alpha}^{\rm{NNA}}+\mu_{\alpha\alpha}^{\rm{NNB}}$ = 0. This is because the nearest-neighbor interaction $J_1$ is nearly zero at ambient pressure and the spin system can be approximated by decoupled two square lattices interacting only via $J_2$. Even though the nearest-neighbor correlations may develop at low temperatures via order by disorder mechanism, such correlations should be of the CAF type, therefore, their sum over all bonds will vanish. Then Eqs.~(\ref{T1c_pc}) and (\ref{T1a_pc}) are simplified as 
\begin{equation}
\begin{split}{\label{T1a_pc2}}
\left(\frac{1}{T_1}\right)_{ab}&=\left(\Gamma_1+\Gamma_3\right)\mu_{\parallel }^{\rm{auto}}
+\left(\Gamma_2+2\Gamma_3\right)\mu_{\perp}^{\rm{auto}} \\
&+\left(\Gamma_1-\Gamma_3\right)\mu_{\parallel}^{\rm{NNN}}
+\left(\Gamma_2-2\Gamma_3\right)\mu_{\perp}^{\rm{NNN}},
\end{split}
\end{equation}
\begin{equation}\label{T1c_pc2}
\left(\frac{1}{T_1}\right)_c=2\left(\Gamma_3\mu_{\parallel }^{\rm{auto}}+\Gamma_2\mu_{\perp}^{\rm{auto}}-\Gamma_3\mu_{\parallel}^{\rm{NNN}}+\Gamma_2\mu_{\perp}^{\rm{NNN}}\right),
\end{equation}
where $\Gamma_1=2\gamma^2_{\rm{N}}A^2_{cc}$, $\Gamma_2=\gamma^2_{\rm{N}}\left(A^2_{aa}+A^2_{bb}\right)$, and $\Gamma_3=\gamma^2_{\rm{N}}A^2_{ac}$. 

Next, we estimate the magnitudes of $\Gamma_1$-$\Gamma_3$. 
The value $A_{cc}$=$-$0.102 T/$\mu_{\rm{B}}$ at ambient pressure (Table~\ref{tab:table1})
gives $\Gamma_1/\gamma^2_{\rm{N}}$=0.021~(T/$\mu_{\rm{B}})^2$. Concerning the hyperfine coupling constants in the $ab$-plane, only the axially symmetric component is determined from the $K$-$\chi$ analysis as $A_{aa}+A_{bb}$=$-$0.054 T/$\mu_{\rm{B}}$ (Table~\ref{tab:table1}). We tentatively assign the limit of in-plane anisotropy as $-0.054 \le A_{aa}, A_{bb} \le 0$. This leads to the possible range of $\Gamma_2$ as $0.00146 \le \Gamma_2/\gamma^2_{\rm{N}} \le 0.00292$~(T/$\mu_{\rm{B}})^2$. The off-diagonal coupling constant $A_{ca}$ depends on the direction $\theta$ of the CAF moment [Eq.~(\ref{A_cac})]. We assume $-\pi/2 \le \theta \le 0$, providing the possible range of $\Gamma_3$ as $1.0 \times 10^{-4} \le \Gamma_3/\gamma^2_{\rm{N}} \le 2.0 \times 10^{-4}~$(T/$\mu_{\rm{B}})^2$. 

There are large differences in the magnitudes of $\Gamma_1$-$\Gamma_3$, $\Gamma_1/\Gamma_2 = 7 \sim 14$ and $\Gamma_2/\Gamma_3 = 7 \sim 29$. Therefore, we may omit the terms proportional to $\Gamma_3$ unless the temperature is very close to $T_{\rm{N}}$, where $\mu_{\alpha \alpha}^{\rm{auto}}=-\mu_{\alpha \alpha}^{\rm{NNN}}$ and all other terms are canceled. Finally we obtain 
\begin{equation}\label{T1a_pc3}
\left(\frac{1}{T_1}\right)_{ab} \sim \Gamma_1\left(\mu_{\parallel }^{\rm{auto}}+\mu_{\parallel}^{\rm{NNN}}\right )+\Gamma_2\left(\mu_{\perp}^{\rm{auto}}+\mu_{\perp}^{\rm{NNN}}\right),
\end{equation}
\begin{equation}\label{T1c_pc2}
\left(\frac{1}{T_1}\right)_c \sim 2\Gamma_2\left(\mu_{\perp}^{\rm{auto}}+\mu_{\perp}^{\rm{NNN}}\right).
\end{equation}

Equation~(\ref{T1c_pc2}) for $(1/T_1)_{c}$ describes general behavior of the relaxation rate of nuclei equally coupled to the two nearest neighbor spins in a square lattice antiferromagnet. A typical example is the oxygen nuclei in the parent compounds of high-$T_c$ cuprates. The results of high temperature series expansion on the 2D Heisenberg model~\cite{Singh1, Gelfand} reveal that $1/T_1$ is constant at high temperatures down to $\sim2J$, below which $1/T_1$ decreases progressively as the pair correlations develop. The experimental result of $(1/T_1)_{c}$ in Fig.~\ref{fig:T1_c_ambient} is consistent with this behavior below 200~K. However, the steep increase of $(1/T_1)_{c}$ above 200~K is quite unusual and difficult to understand. 

A even more puzzling result is the nonmonotonic temperature dependence of the anisotropy $R$ of correlation functions 
\begin{equation}\label{R}
R=\frac{\mu_{\parallel }^{\rm{auto}}+\mu_{\parallel}^{\rm{NNN}}}{\mu_{\perp}^{\rm{auto}}+\mu_{\perp}^{\rm{NNN}}},
\end{equation}  
which can be directly obtained from the anisotropic ratio of $1/T_1$, 
\begin{equation}\label{T1_ratio}
\frac{(1/T_1)_{ab}}{(1/T_1)_{c}} = \frac{1}{2} + \frac{\Gamma_1}{2\Gamma_2}R.
\end{equation}  
The plot in Fig.~\ref{fig:T1_ratio_ambient} shows the temperature dependence of the anisotropic ratio $(1/T_1)_{ab}/(1/T_1)_{c}$ with the left axis. The right axis indicates the value of $R$ obtained from Eq.~(\ref{T1_ratio}) by tentatively assuming the lower limit of $\Gamma_2$ as $\Gamma_2/\gamma^2_{\rm{N}}$ = 0.00146~(T/$\mu_{\rm{B}})^2$. At 300~K, $R \sim$ 0.2, indicating strong easy plane anisotropy. With lowering temperature down to 100~K, $R$ increases by a factor of more than three, then stays nearly constant down to 50~K. With further lowering temperature, $R$ is reduced again to small values.

\end{document}